\documentclass[twocolumn,showpacs,preprintnumbers,amsmath,amssymb]{revtex4-1}

\usepackage{graphicx}
\usepackage{dcolumn}
\usepackage{bm}
\begin{document}
\preprint{APS/123-QED}
\title{In-medium $\omega$ mass from the $\gamma + Nb \rightarrow
  \pi^{0}\gamma + X$ reaction}
\author{~M.~Nanova$^{1}$,~V.~Metag$^{1}$, G. Anton$^{2}$, J.C.S. Bacelar$^{3}$, O. Bartholomy~$^4$, D.~Bayadilov$^{4,5}$, Y.A.~Beloglazov$^5$, R.~Bogend\"orfer$^2$,~R.~Castelijns$^{3}$, V.~Crede$^{4,a}$, H.~Dutz$^6$, A.~Ehmanns$^4$, D.~Elsner$^6$, K.~Essig$^4$, R.~Ewald$^6$,~I.~Fabry$^4$, M.~Fuchs$^4$, Ch.~Funke$^4$, R.~Gothe$^{6,b}$, R.~Gregor$^1$, A.B.~Gridnev$^5$, E.~Gutz$^4$, S.~H\"offgen$^4$,~P.~Hoffmeister$^4$, I.~Horn$^4$, J. H\"ossl$^2$, I.~Jaegle$^7$, J.~Junkersfeld$^4$, H.~Kalinowsky$^4$,~Frank~Klein$^6$,~Friedrich~Klein$^6$, E.~Klempt$^4$, M.~Konrad$^6$, B.~Kopf$^{8,9}$, M.~Kotulla$^1$, B.~Krusche$^7$, J.~Langheinrich$^{6,9}$, H.~L\"ohner$^3$, I.V.~Lopatin$^5$, J.~Lotz$^4$, S.~Lugert$^1$, D.~Menze$^6$, T.~Mertens$^7$, J.G. Messchendorp$^3$, C.~Morales$^6$,~R.~Novotny$^1$, M.~Ostrick$^{6,c}$, L.M.~Pant$^{1,d}$, H.~van Pee$^4$, M.~Pfeiffer$^1$, A. Roy$^{1,e}$, A.~Radkov$^5$,~S.~Schadmand$^{1,f}$, Ch.~Schmidt$^4$, H.~Schmieden$^6$, B.~Schoch$^6$, S.~Shende$^{3,i}$, G. Suft$^2$,~A.~S\"ule$^6$, V.~V.~Sumachev$^5$, T.~Szczepanek~$^4$, U.~Thoma$^4$, D.~Trnka$^1$, R.~Varma$^{1,g}$, D.~Walther$^4$, Ch.~Weinheimer$^{4,h}$,  Ch.~Wendel$^4$\\
  (The CBELSA/TAPS Collaboration)}
\affiliation{%
{$^{1}$II. Physikalisches Institut, Universit\"at Gie{\ss}en, Germany}\\
{$^{2}$Physikalisches Institut, Universit\"at Erlangen, Germany}\\
{$^{3}$Kernfysisch Versneller Institut  Groningen, The Netherlands}\\
{$^{4}$Helmholtz-Institut f\"ur Strahlen- u. Kernphysik Universit\"at Bonn, Germany }\\
{$^5$Petersburg Nuclear Physics Institute Gatchina, Russia}\\
{$^{6}$Physikalisches Institut, Universit\"at Bonn, Germany}\\
{$^{7}$Physikalisches Institut, Universit\"at Basel, Switzerland}\\
{$^{8}$Institut f\"ur Kern- und Teilchenphysik TU Dresden, Germany}\\
{$^{9}$Physikalisches Institut, Universit\"at Bochum, Germany}\\
{$^a$Current address: Florida State University Tallahassee, FL, USA}\\
{$^b$Current address: University of South Carolina Columbia, SC, USA}\\
{$^c$Current address: Physikalisches Institut Universit\"at Mainz, Germany}\\
{$^d$Current address: Nuclear Physics Division BARC, Mumbai, India}\\
{$^e$Current address: IIT Indore, Indore, India}\\
{$^f$Current address: Forschunszentrum J\"ulich, Germany}\\
{$^g$Current address: Department of Physics I.I.T. Powai, Mumbai, India}\\
{$^h$Present address: Universit\"at M\"unster, Germany}            \\
{$^i$Current address: Department of Physics, Univ. of Pune, Pune, India}\\
}
\date{\today}
\begin{abstract}
Data on the photoproduction of $\omega$ mesons on nuclei have been re-analyzed
in a search for in-medium modifications. The data were taken with the
Crystal Barrel(CB)/TAPS detector system at the ELSA accelerator facility in
Bonn. First results from the analysis of the data set were published by
D. Trnka et al. in Phys. Rev. Lett 94 (2005) 192303  \cite{david}, claiming a
lowering of the $\omega$ mass in the nuclear medium by 14$\%$ at normal nuclear
matter density. The  extracted $\omega$ line shape was found to be sensitive to
the background subtraction. For this reason a re-analysis of the same data set
has been initiated and a new method has been developed to reduce the background
and to determine the shape and absolute magnitude of the background directly
from the data. Details of the re-analysis and of the background determination
are described. The $\omega$ signal on the $Nb$ target, extracted in the
re-analysis, does not show a deviation from the corresponding line shape on a
$LH_2$ target, measured as reference. The earlier claim of an in-medium mass
shift is thus not confirmed. The sensitivity of the $\omega$ line shape to
different in-medium modification scenarios is discussed. 
\end{abstract}
\maketitle
\section{Introduction}
Quantum Chromodynamics (QCD) has been remarkably successful in describing
strong interactions at high energies ( $\gg 10$ GeV ) and short distances
($\leq 10^{-2}$ fm) where quarks and gluons are the relevant degrees of
freedom. At these scales the strong coupling is so small ($\alpha_s \approx
0.1$) that perturbative treatments provide a first order description of the
phenomena \cite{Gross,Politzer,Wilczek}. Applying QCD at lower energies is a
major challenge. In the GeV energy range the coupling strength among quarks
and gluons becomes very large and hadrons - composite objects made of quarks
and gluons - emerge as the relevant degrees of freedom. A rigorous way to
solve QCD in this energy regime is lattice QCD. With the advent of high speed
supercomputers remarkable progress has been achieved in lattice QCD
simulations with dynamical $u,d,$ and $s$ quarks. D\"urr et al \cite{Duerr}
have recently succeeded in reproducing masses of mesons and baryons within
3$\%$ of the experimental values.\\ 
While the properties of free hadrons are in
most cases experimentally known with resasonable accuracy a possible
modification of these properties in a strongly interacting medium is a much
debated issue. In fact, in-medium changes of hadron properties have been
identified as one of the key problems in understanding the non-perturbative
sector of QCD. Fundamental symmetries in QCD provide guidance in dealing with
strong interaction phenomena in the non-perturbative domain. Furthermore, QCD
sum rules have been applied to connect the quark-gluon sector to hadronic
descriptions. Along these lines, QCD inspired hadronic models have been
developed to calculate the in-medium self-energies of hadrons and their
spectral functions. Mass shifts and/or in-medium broadening as well as more
complex structures in the spectral function due to the coupling of vector
mesons to nucleon resonances have been predicted. A recent overview is given
in \cite{RWH}. These studies have motivated widespread experimental attempts
to confirm or refute these theoretical predictions.\\ 
Heavy-ion collisions and reactions with photons and protons have been used to
extract experimental information on in-medium properties of hadrons. The
experiments have focused on the light vector mesons $\rho, \omega$ and $\phi$ since their decay lengths are  comparable to nuclear dimensions after being produced in some nuclear reaction. To ensure a reasonable decay probability in the strongly interacting medium cuts on the recoil momentum are, however, required for the longer lived $\omega$ and $\phi$ mesons.\\
A full consensus has not yet been reached among the different
experiments. A detailed account of the current status of the field is given in
comprehensive reviews \cite{Hayano_Hatsuda,LMM}. An in-medium broadening of
the vector mesons is reported by almost all experiments and the majority of
experiments does not find evidence for a mass shift. Apart from \cite{david}
only one other experiment at KEK \cite{Naruki} reports a drop of the $\rho$
and $\omega$ mass by 9 $\%$ at normal nuclear matter density. Studying
$\omega$ meson production in ultra-relativistic heavy-ion collisions, the NA60
collaboration observes a suppression of the meson yield for $\omega$ momenta
below 1 GeV/c which is even more pronounced for more central collisions
\cite{Arnaldi}. This is interpreted as evidence for in-medium modifications of
slow $\omega$ mesons but it cannot be concluded whether this is due to a mass
shift, a broadening, or both.\\ 
It should be noted that a search for mass shifts has turned out to be much
more complicated than initially thought for those cases where a strong
broadening of the meson is observed as for the $\omega$ \cite{Kotulla} and
$\phi$ meson \cite{Ishikawa}. In the $\omega \rightarrow \pi^0 \gamma$ decay mode the increase in the total width of $\omega$ drastically lowers the branching ratio for in-medium decays into this  channel and thereby reduces the sensitivity of the observed $\omega$ signal to in-medium modifications.\\  
In this paper data on the photoproduction of $\omega$ mesons on $Nb$ and
$LH_2$ are re-analyzed which were taken with the CB/TAPS detector system
at the ELSA accelerator facility in Bonn. First results from an analysis of
these data were published by D. Trnka et al. \cite{david}, claiming a mass
shift of the $\omega$ meson by -14$\%$ at normal nuclear matter density. This
information was extracted from a comparison of the $\omega$ signals on $Nb$ and
$LH_2$, reconstructed in the $\pi^0 \gamma$ channel. As pointed out in the
literature \cite{Kaskulov} the deduced line shapes are very sensitive to the
background subtraction. While in the initial work the background was
determined by fitting the $\pi^0 \gamma $ invariant mass spectrum a much more
refined background determination is used in the current analysis. The paper
gives a full account of the experiment and details of the analysis steps.   
\begin{figure*} 
 \resizebox{0.9\textwidth}{!}{
     \includegraphics[height=.3\textheight]{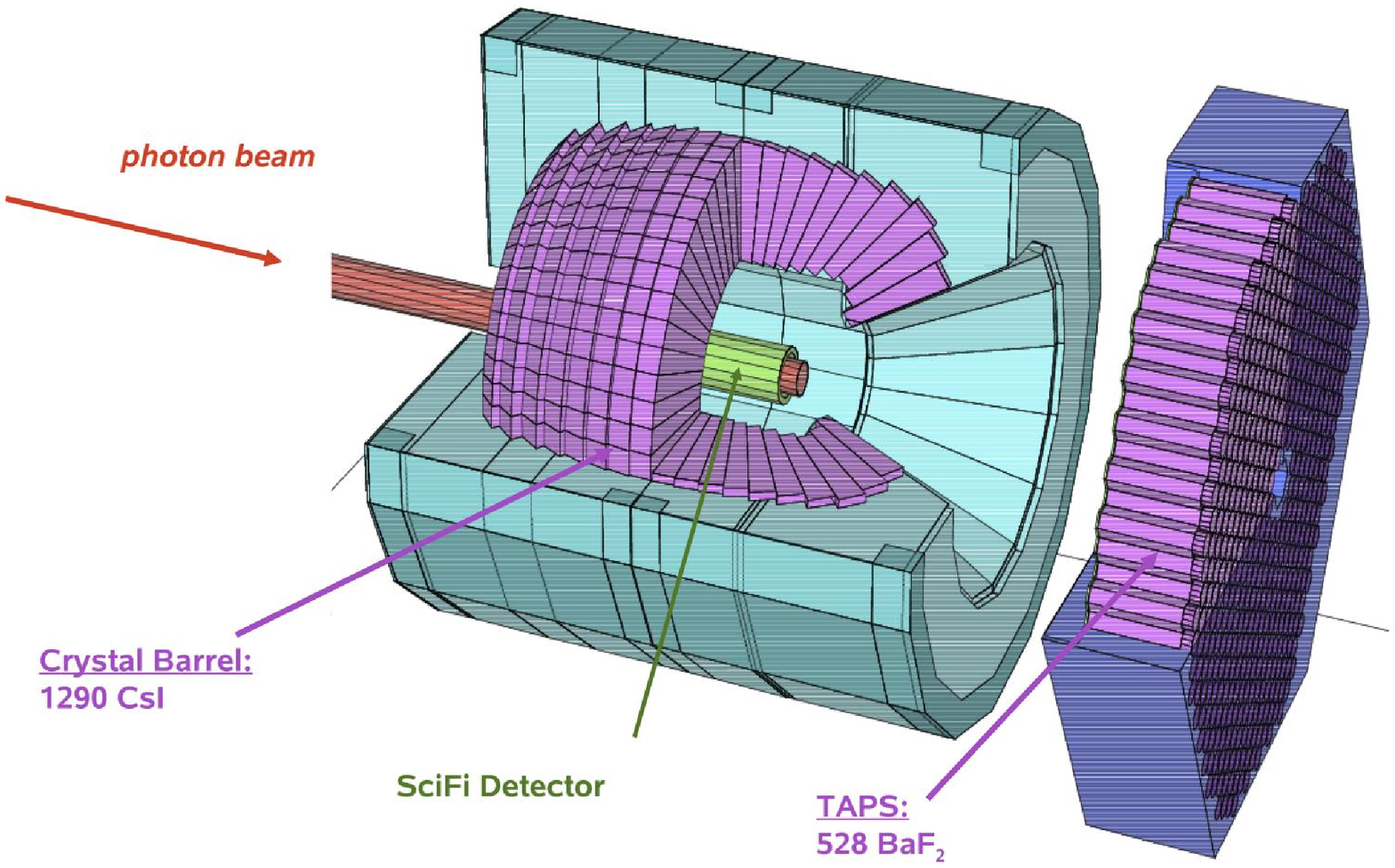}
      \includegraphics[height=.28\textheight]{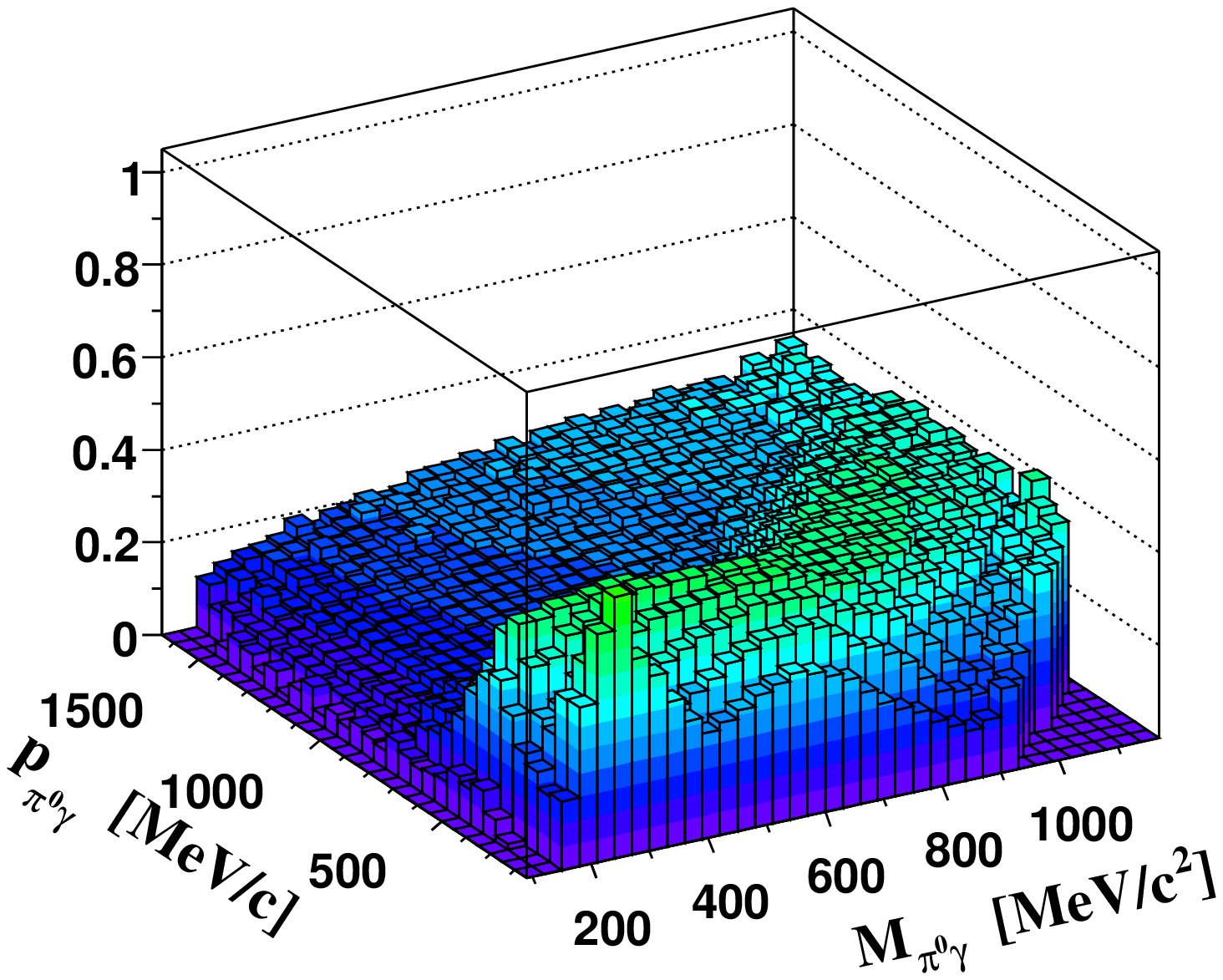}}
 \caption{(Color online) Left: Sketch of the  CB/TAPS setup. The tagged photons impinge      on the nuclear target in the center of the Crystal Barrel detector. The   TAPS detector at a distance of 1.18 m from the target serves as a forward wall of the Crystal Barrel. The combined detector system provides photon detection capability over almost the full solid angle. Charged particles leaving the target are identified in the inner scintillating-fiber detector and in the plastic scintillators in front of each BaF$_2$ crystal in TAPS. Right: Detector acceptance for the $ p \pi^{0} \gamma$ final state as a function of the invariant mass and momentum of the $\pi^0 \gamma$ pair for incident photon energies of 900 to 2200 MeV.} \label{fig:exp} 
\end{figure*}

\section{\label{sec:level1}{Experimental Setup}}
\subsection{\label{sec:level2}CB/TAPS detector system at ELSA}
Data on $LH_2 , C, $ and $Nb$ have been taken with the detector system Crystal
Barrel (CB)~\cite{aker92} and TAPS~\cite{Novotny1,Gabler1} at the electron
stretcher facility ELSA~\cite{Husmann,Hillert}. The detector setup is shown
schematically in Fig.~\ref{fig:exp} left. Electrons extracted from ELSA with energy
$E_0$ hit a primary radiation target, a thin copper or diamond crystal, and
produce bremsstrahlung ~\cite{Elsner}. The energy of the bremsstrahlung
photons is determined eventwise from the deflection of the scattered electrons
in a magnetic field. The detector system in the focal plane of the magnet
consists of 480 scintillating fibers and 14 partly overlapping scintillator
bars. From the energy of the scattered electron $E_e^{-}$ the energy of the
photon impinging on the nuclear target is given by $E_\gamma = E_0 -
E_e^{-}$. Photons were tagged in the energy range from 0.5 GeV up to 2.6 GeV
for an incoming electron energy of 2.8 GeV. The total tagged photon intensity
was about $10^{7}$ $s^{-1}$ in this energy range. The energy resolution varied
between 2 MeV for the high photon energies and 25 MeV for the low photon
energies, respectively. The part of the beam that did not produce any
bremsstrahlung photons was deflected by the magnet as well. Since these
electrons retained their full energy the curvature of their track is smaller
and they passed over the tagger into a beam dump.\\ 
The Crystal Barrel (CB) detector, a photon calorimeter consisting of 1290
CsI(Tl) crystals ($\approx$16 radiation lengths), covered the complete
azimuthal angle and the polar angle from $30^o$ to $168^o$. The $LH_2 , C $
and $Nb$ targets (30 mm in diameter, 53 mm, 20 mm and 1 mm thick,
respectively) were mounted in the center of the CB, surrounded by a
scintillating fibre-detector to register charged particles ~\cite{suft}. The
CB was combined with a forward detector - the TAPS calorimeter - consisting of
528 hexagonal BaF$_2$ crystals ($\approx$12 $X_0$), covering polar angles
between $5^o$ and $30^o$ and the complete azimuthal angle. In front of each
BaF$_2$ module a 5 mm thick plastic scintillator was mounted for the
identification of charged particles. The combined CB/TAPS detector covered
99\% of the full 4$\pi$ solid angle. The high granularity of this system makes
it very well suited for the detection of multi-photon final states.\\

\subsection{\label{sec:level3}The trigger}
$\omega$ mesons produced by photons on a nuclear target were identified via their $\omega \rightarrow \pi^0 \gamma \rightarrow \gamma \gamma \gamma$ decay. Events with $\omega$ candidates (3 photons in the final state) were selected with suitable trigger conditions: the first level trigger was derived from TAPS, requiring either $\geq$2 hits
above a low threshold (A) or, alternatively, $\geq$1 hit above a high
threshold (B). The second level trigger (C) was based on a fast cluster
recognition (FACE) logic, providing the number n of clusters in the Crystal
Barrel as well as their energy and angle within $\simeq$10 $\mu$s .
For the data on the solid target the total
trigger condition required $\lbrack A \vee (B\wedge C )\rbrack$, with n = 2 clusters identified on the second level (C). Events with two photon candidates were taken for calibration purposes and for 
checking the analysis with known cross sections.

\subsection{\label{sec:level4}Detector acceptance}
Although the CB/TAPS detector system covers almost the full solid angle it is
nevertheless very important to study the acceptance for reconstructing the
reaction of interest. Monte Carlo (MC) simulations of the reaction $\gamma A
\rightarrow X p \pi^{0} \gamma $ have been performed for solid targets using the
GEANT3 package, assuming a phase space distribution of the final state
particles and taking the Fermi motion of nucleons in the target nucleus into
account. The reconstruction of simulated $\pi^{0} \gamma$ data is done for the
same trigger conditions as in the experiment and for the incident photon
energy range from 900 to 2200 MeV. The acceptance as a function of the
invariant mass and the momentum of the $p \pi^0 \gamma$ final state is shown in
Fig.~\ref{fig:exp} right.  In the $\omega$ mass range the acceptance is rather flat as a function of momentum and amounts to $\approx$ 35$\%$.
\begin{figure*} 
  \resizebox{0.8\textwidth}{!}{%
    \includegraphics{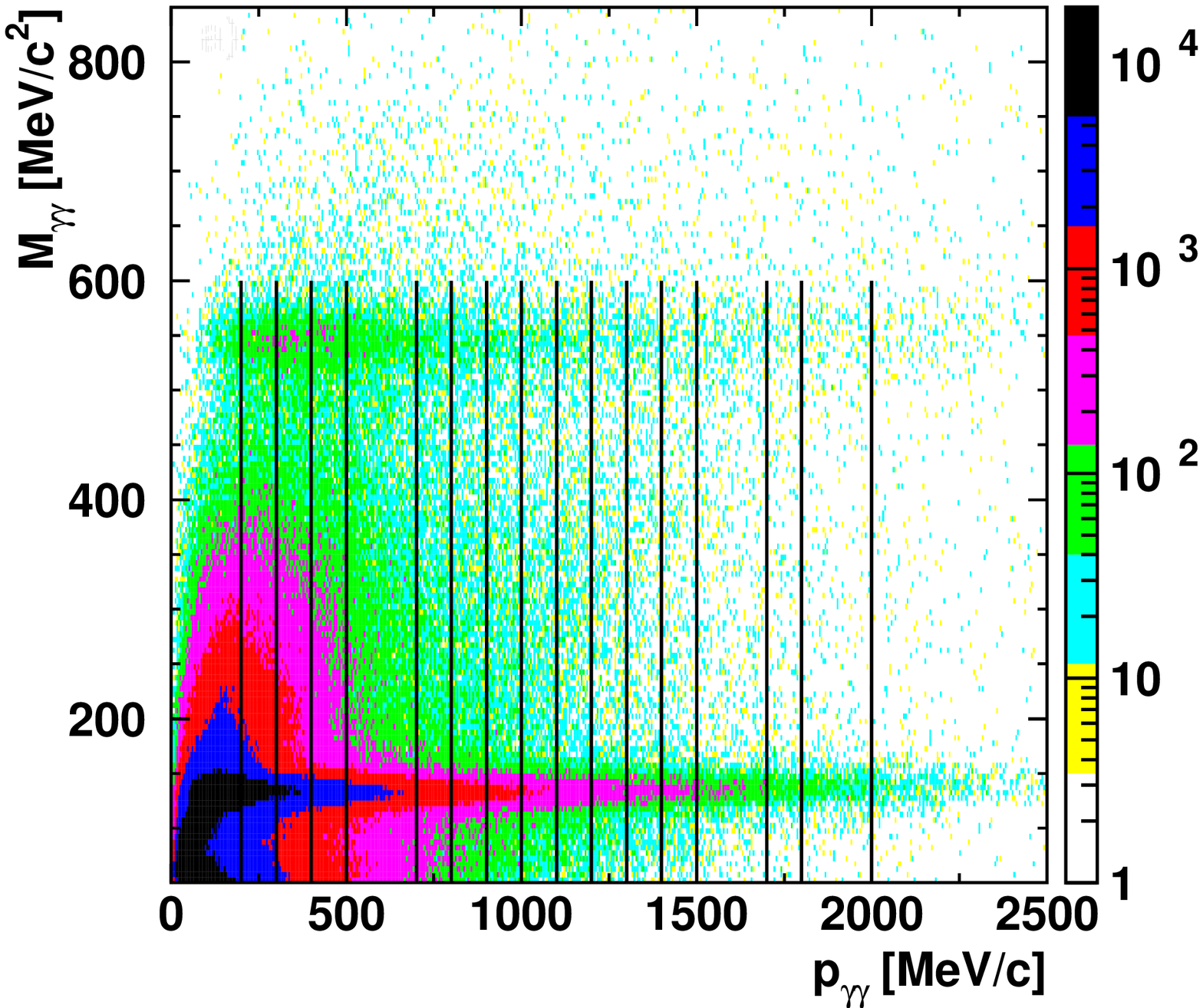}\includegraphics{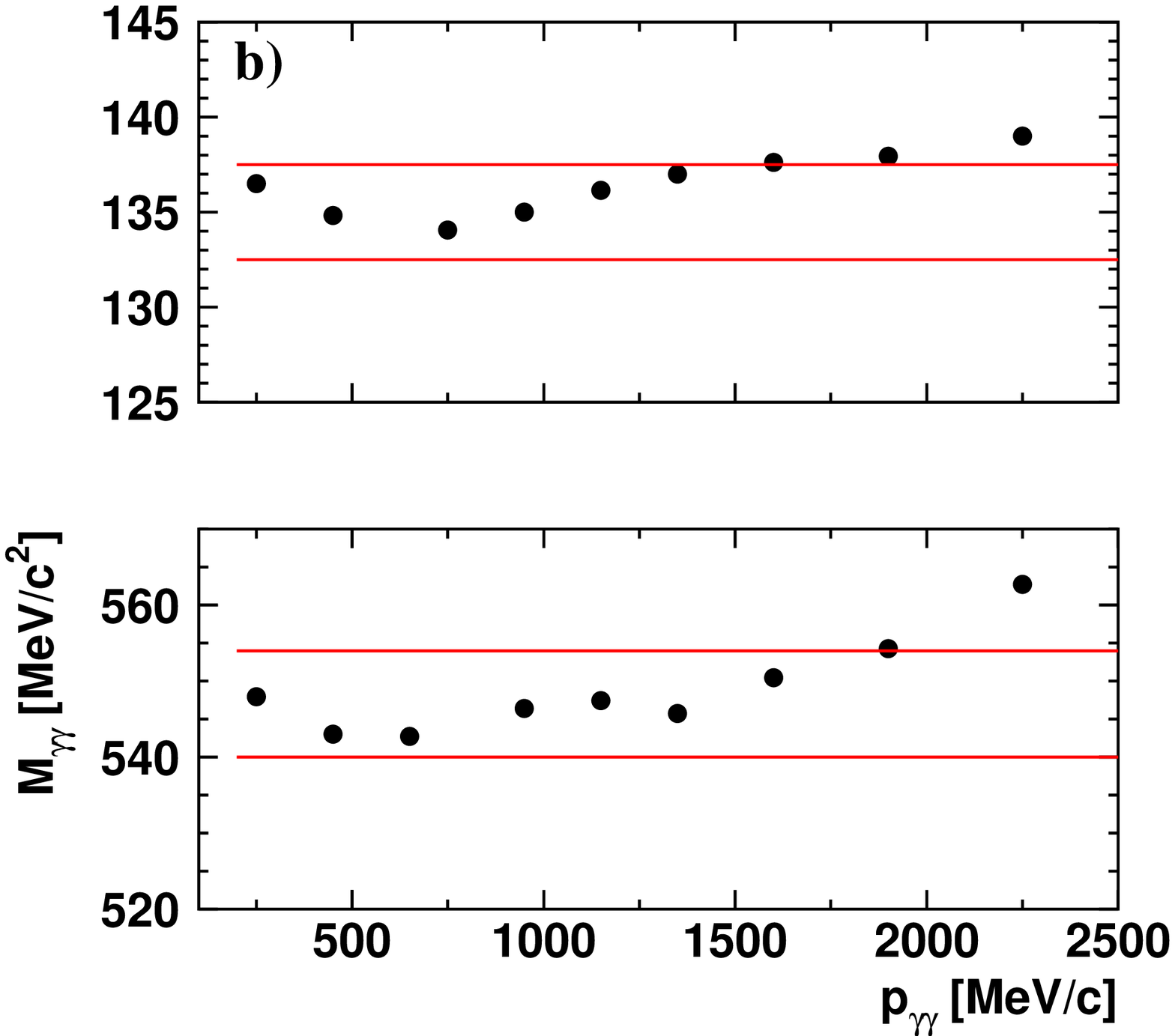}}     
  \resizebox{0.8\textwidth}{!}{%
    \includegraphics{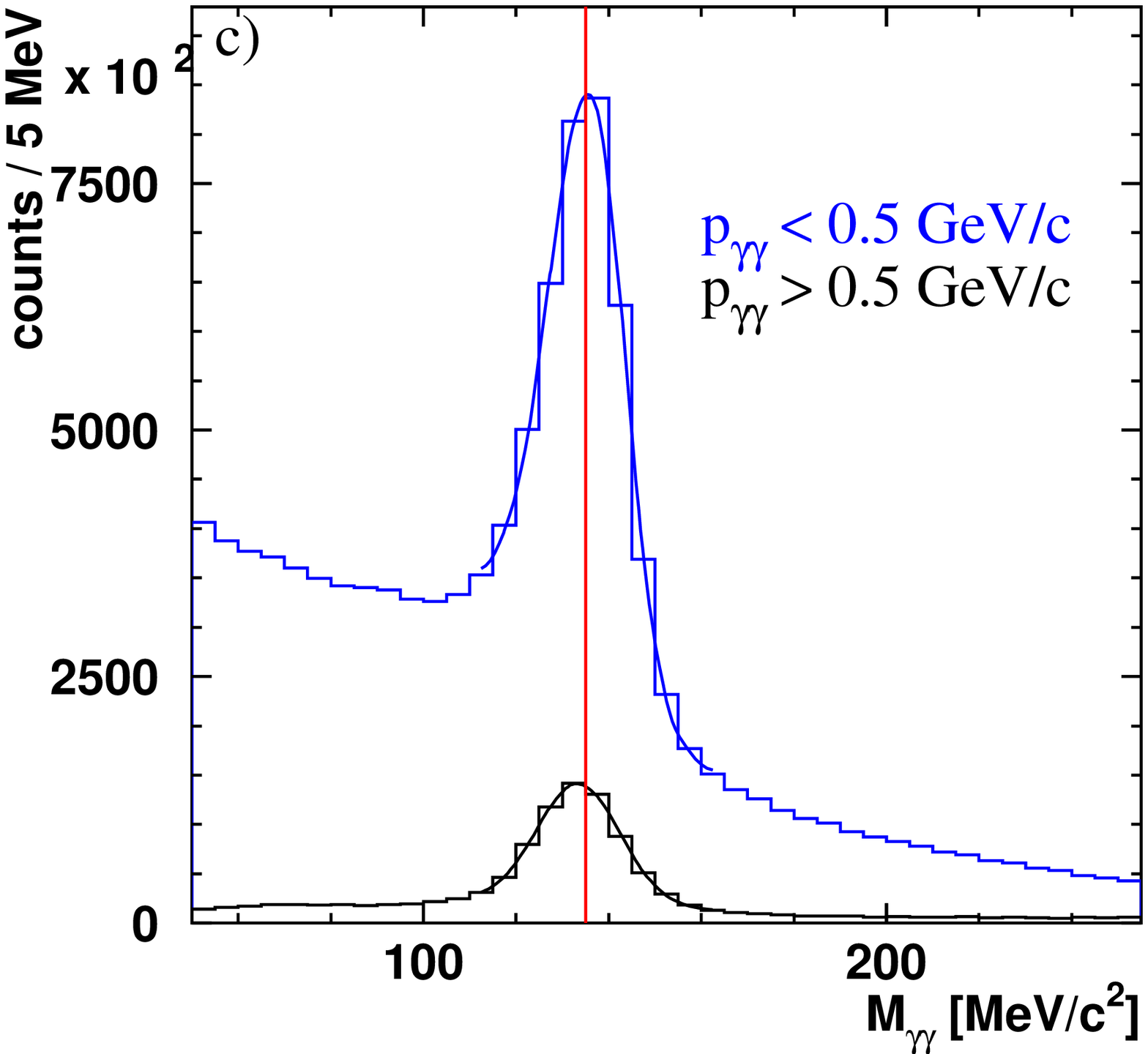}\includegraphics{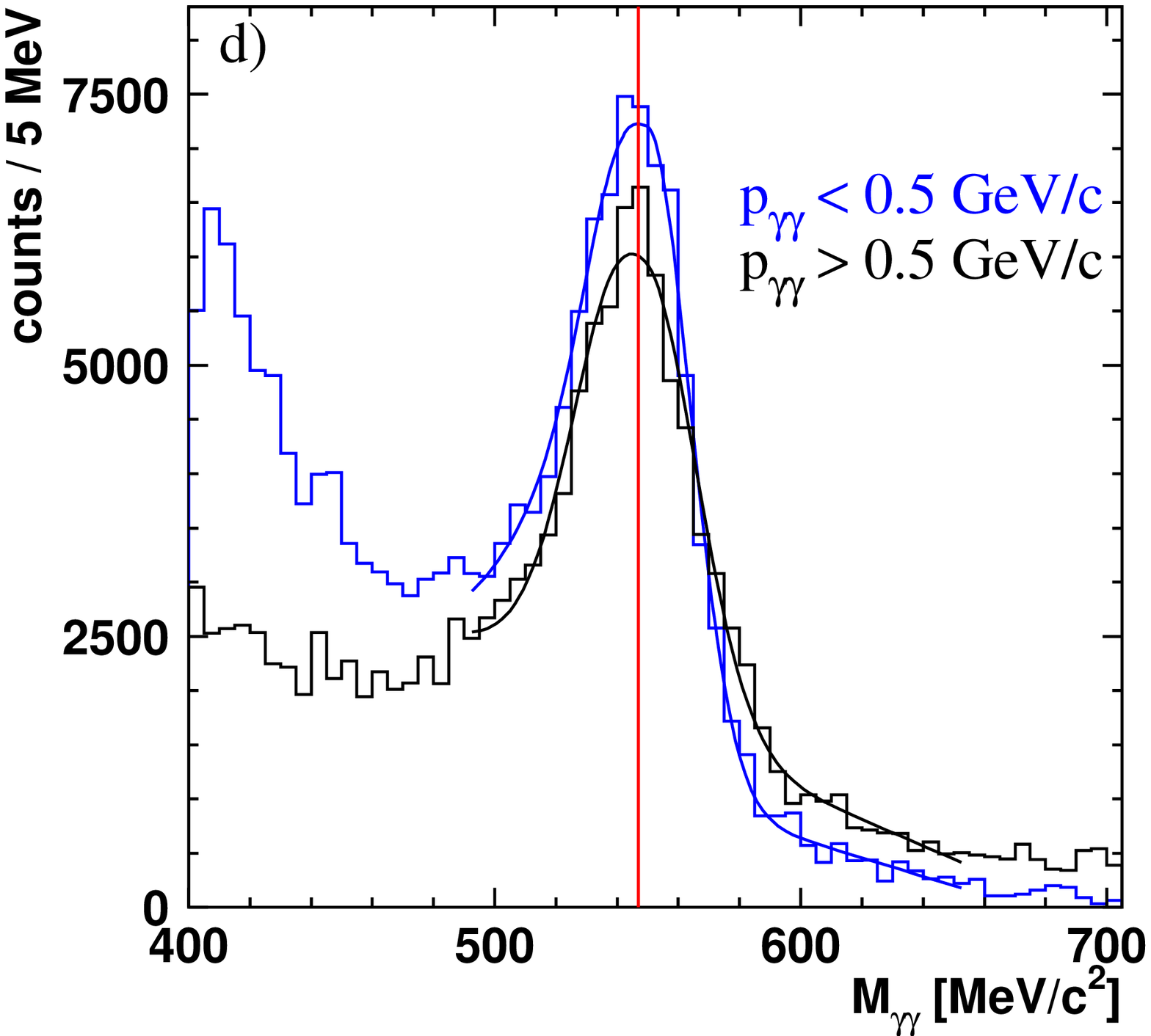}}
\caption{(Color online) a) Invariant mass of two $\gamma$'s ($\pi^{0} \rightarrow \gamma \gamma$ and $\eta \rightarrow \gamma \gamma$) as a function of the momentum of the 2$\gamma$ pair for the $Nb$ target. The vertical lines show the slices for the projections on the y-axis. b) The peak position of the $\pi^{0}$ and $\eta$ invariant mass in 8 slices of the momentum. The horizontal lines show the tolerance of $\pm$2.5 MeV of the $\pi^0$ mass ($135 MeV/c^2$) and of $\pm$7 MeV of the $\eta$ mass ($547 MeV/c^2$). c) The $\pi^{0}$ invariant mass for low and high momenta. d) The $\eta$ invariant mass for low and high momenta.} \label{fig:calib}
\end{figure*}
\begin{figure*} 
  \resizebox{0.9\textwidth}{!}{
     \includegraphics{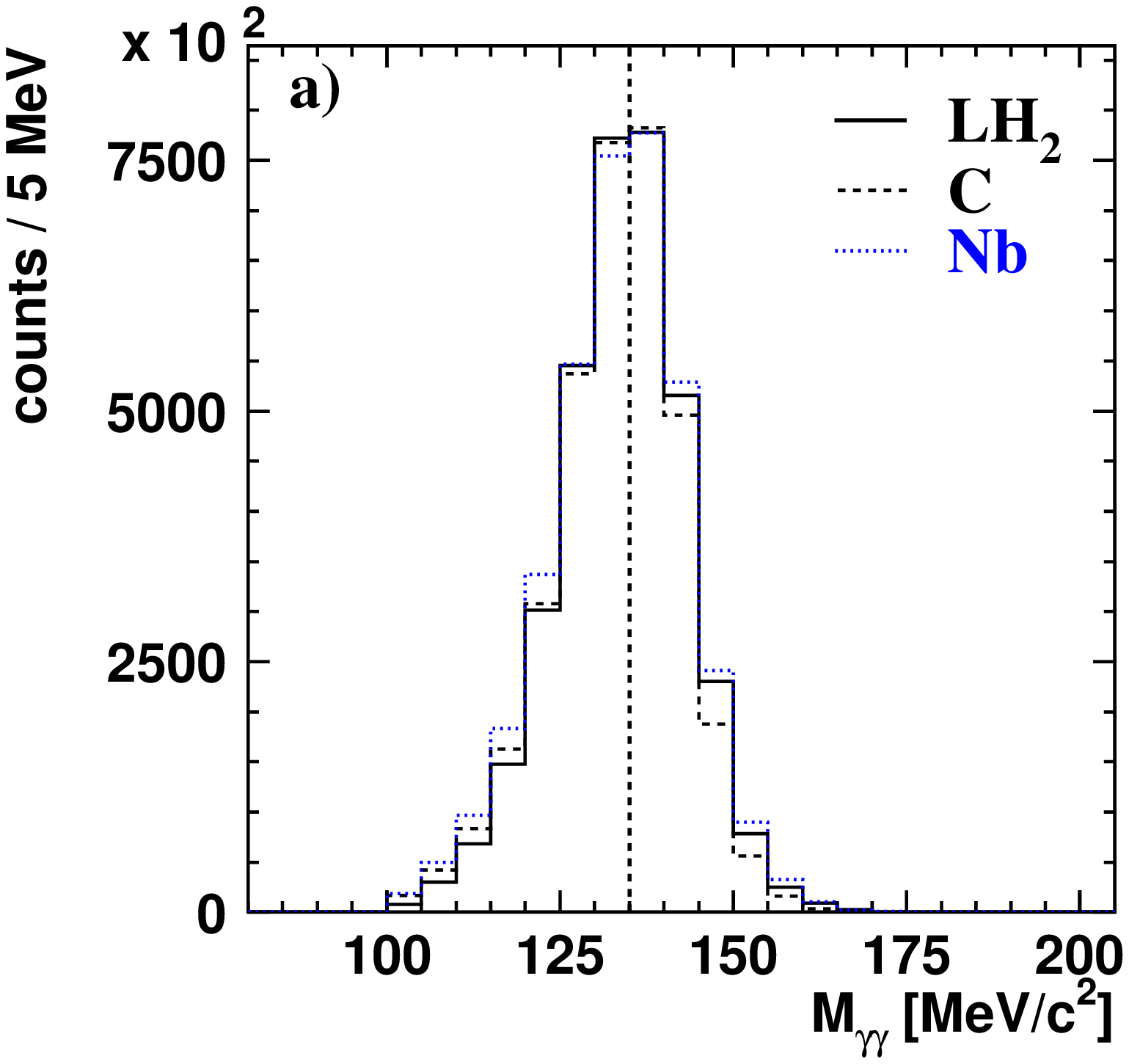}\includegraphics{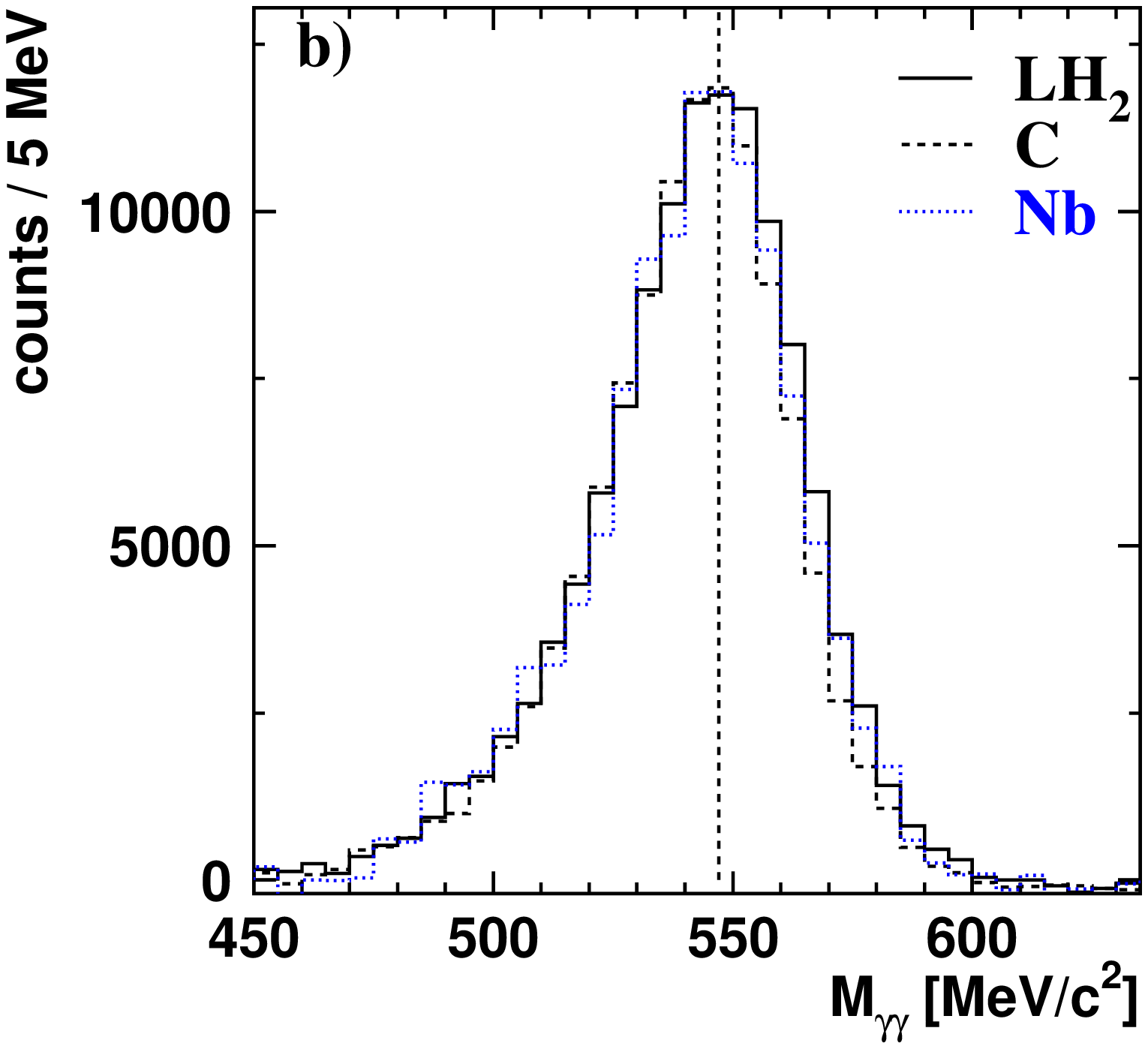}}
   \caption{(Color online) $\pi^0$ (a) and $\eta$ (b) invariant mass distributions
    reconstructed from the $\pi^0 (\eta) \rightarrow \gamma \gamma$ decay for
    the $LH_2, C$ and $Nb$ targets after background subtraction.} \label{fig:calib_2}
\end{figure*} 

\maketitle
\section{\label{sec:level5}{Analysis}}
\subsection{\label{sec:level6}Calibration}
 
\begin{figure*} 
  \resizebox{1.\textwidth}{!}{
    \includegraphics{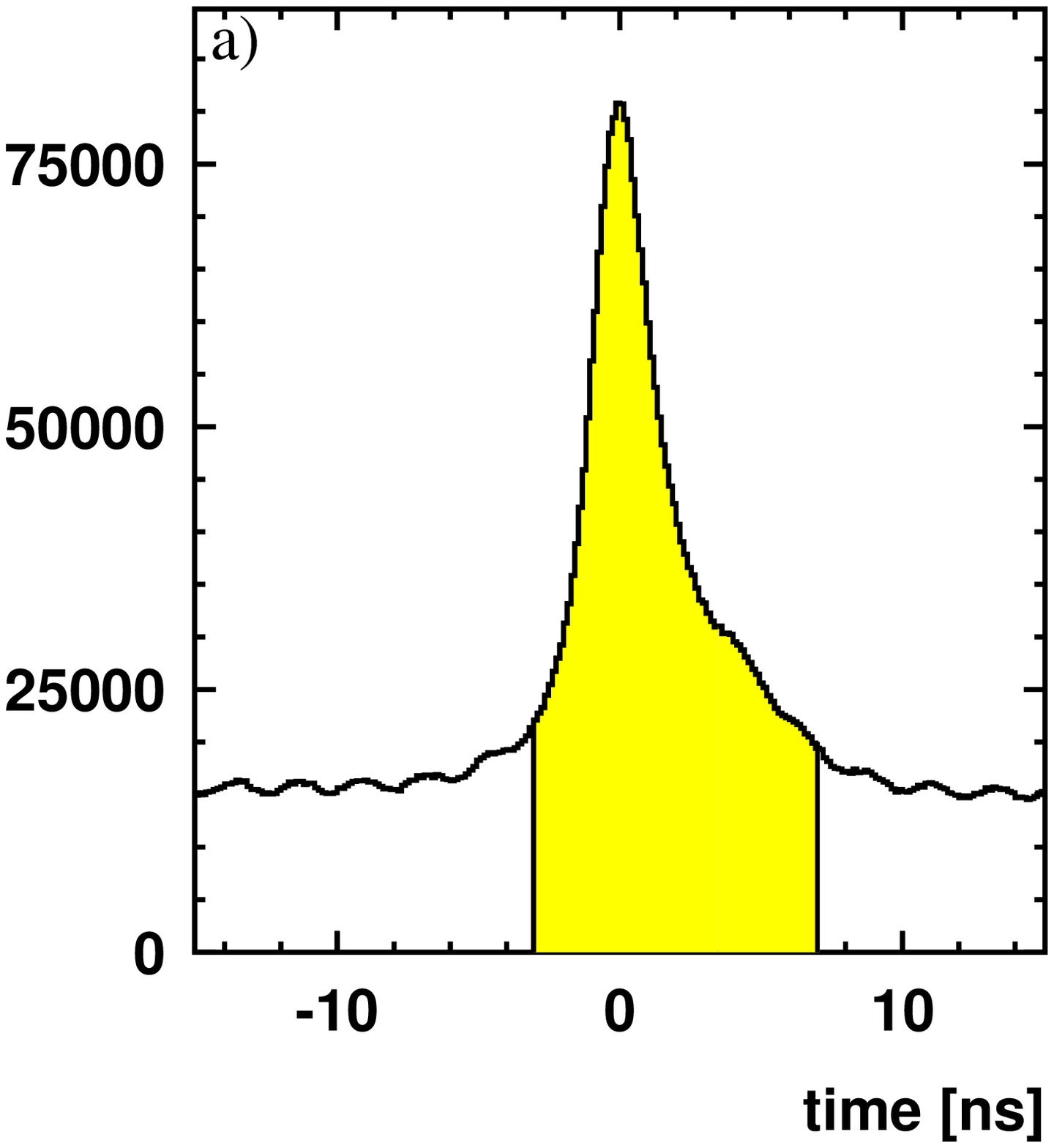} \includegraphics{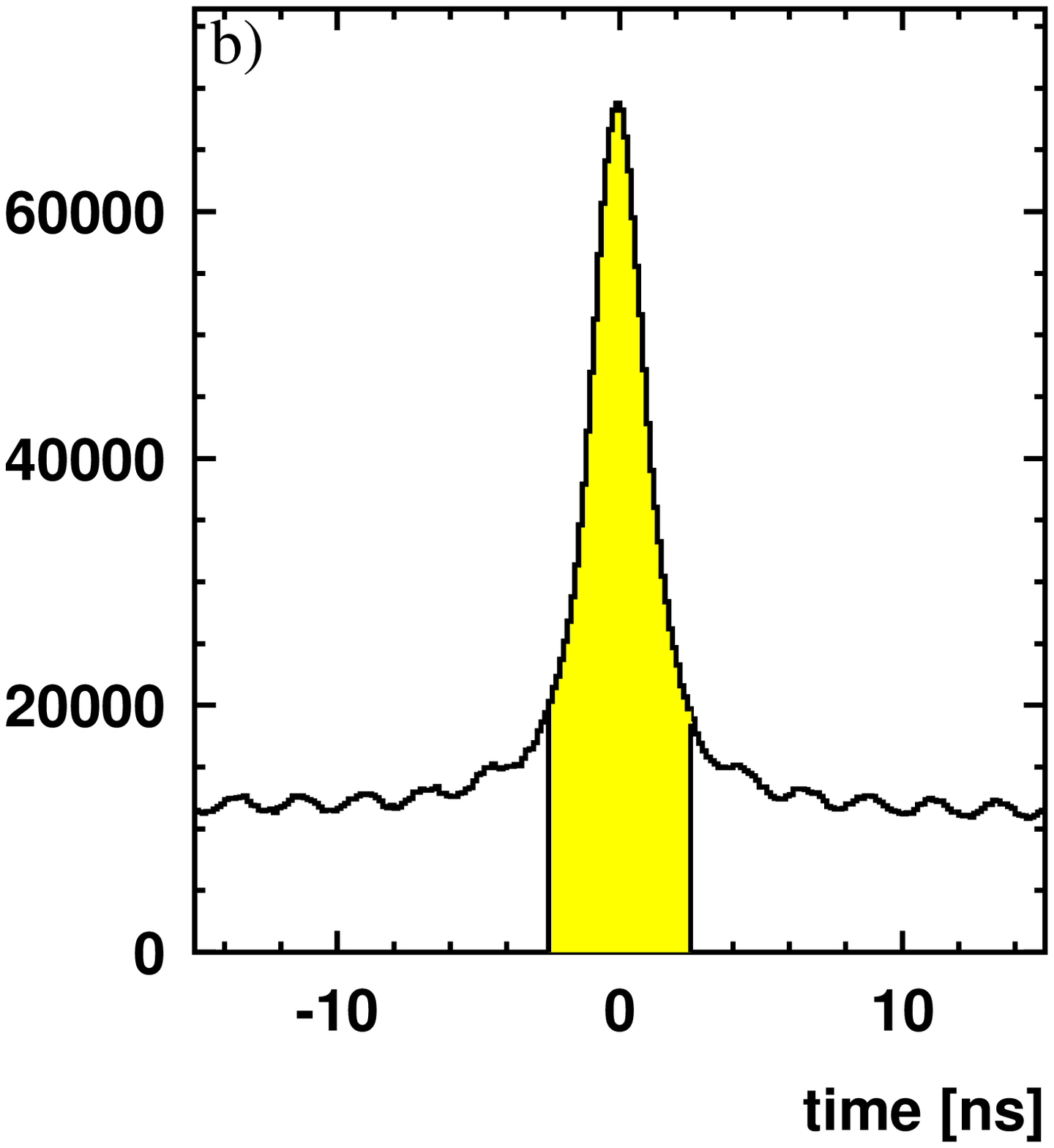}}
\caption{(Color online) TAPS-tagger coincidence time spectra without (a) and with (b)
    requiring the hit in TAPS to be due to a photon (no response in plastic scintillator in front of TAPS). The shaded areas represent the applied cuts. The peaks reside on an uniformly distributed
    background stemming from random coincidences.}\label{fig:time}
\end{figure*}
Since the experiment searches for possibly small mass shifts it is absolutely
mandatory to verify the linearity, accuracy and stability of the photon energy
calibration. The accurately known masses of the $\pi^0$ and $\eta$ meson are
used as calibration fix points since the decay photons of $\pi^0$ and $\eta$ mesons cover the full range in energy of the $\omega$ decay photons. The invariant masses of the mesons were
calculated from the measured 4 momenta of the decay photons. To ensure the
stability of the photon energy calibration the invariant mass of $\pi^{0}$- and
$\eta$-mesons is checked for different momentum bins. For this check a
2-dimensional plot of the $\pi^{0}$($\eta$) invariant mass against the
momentum $\arrowvert \overrightarrow{p}\arrowvert_{\pi (\eta)}$of the
$\pi^0(\eta)$ is filled (Fig.~\ref{fig:calib}a) and projected onto the $\pi^{0}$($\eta$) invariant
mass axis for different slices in the momentum of the $\gamma \gamma$ pair.
Changes in the $\pi^{0}$ and $\eta$ meson invariant mass with momentum are
found to be less than $\leq$ 1.9 \% and 1.3\%, respectively (Fig.~\ref{fig:calib}b). The peak position of $\pi^{0}$ at 135 MeV and of $\eta$ at 547 MeV is stable for different cuts on
the momentum like $>$ 500 MeV or $<$ 500 MeV (Fig.~\ref{fig:calib}c,d).
In addition, it has been verified that the energy calibrations for the runs
with different targets are in agreement. This is demonstrated in
Fig.~\ref{fig:calib_2} which shows the signal line shapes for the $\pi^0$ and
$\eta$ meson measured via their two photon decays for the $LH_2, C$ and $Nb$ targets.

\subsection{\label{sec:level7}Event Selection}
$\omega$ mesons were reconstructed in the reaction $\gamma A \rightarrow (A-1) p \omega
\rightarrow (A-1) p \pi^{0} \gamma$  from events with 3 photons and one proton in the final
state in contrast to the analysis by D. Trnka et al. \cite{david} where the
fourth particle was not further identified. In a first step only those events
were selected which had 4 hits, so called PED (particle energy deposit), in
the detector system. In order to reconstruct the reaction for $\omega$
photoproduction 1 charged particle was required in coincidence with 3 neutral
hits (from the 4 PED data set) in the CB/TAPS detector system. The selection of
the charged particles was done by using either the information from the fiber
detector in the CB or the information from the plastic scintillators in front
of the TAPS detector. Requesting a charged particle in addition to 3 neutral
hits leads to a loss in statistics, but is essential for the background
determination described in section ~\ref{sec:level16}.\\  
The possible background contributions were investigated via Monte Carlo
simulations. The reactions $\gamma A \rightarrow (A-1) p \pi^{0} \pi^{0}
\rightarrow (A-1) p 4 \gamma$ and $\gamma A \rightarrow (A-1) p \pi^{0} \eta
\rightarrow  (A-1) p 4 \gamma$,
where one of the photons in the final state escaped detection, were found to
be the dominant background sources. Furthermore the reaction $\gamma A
\rightarrow (A-1) n \pi^{+} \pi^{0}$ where the neutron and the $\pi^{+}$ are
misidentified as a photon and a proton, respectively, also contribute to the
background. For the analysis which is presented here the background was
reconstructed from 5 PED events with 4 neutral and 1 charged particle (see
section ~\ref{sec:level16}).\\ 

\subsection{\label{sec:level8}Reconstruction of the $\omega$ meson}
\subsubsection{\label{sec:level9}Incident photon energy range}
The analysis was performed for incident photon energies from 900 to 2200 MeV,
i.e. starting about 200 MeV below the $\omega$ production threshold off the
free nucleon $E^N_{\gamma,thresh}$=1109 MeV. The threshold for $\omega$
production on nuclei is given by the threshold for coherent production
\begin{equation}  
E_{\gamma,thresh} = m_{\omega} + \frac{m^{2}_{\omega}}{2m_{A}}
\label{coh_thresh} 
\end{equation}
where the recoil momentum of the produced meson is taken up by the whole
nucleus. For a $Nb$ target Eq.~\ref{coh_thresh} yields a coherent threshold
energy of $E_{\gamma,thresh}$ = 786 MeV, i.e. the threshold is even lower
than 900 MeV. The choice of the incident energy interval represents a
compromise between sufficiently low energies for $\omega$ production off a
nuclear target and sufficient discrimination of background sources, which
strongly increase with decreasing photon energies; e.g., the 2 $\pi^0$ channel, which is the strongest background channel, exhibits maxima in the cross section for incident photon energies near 1080 and 750 MeV.

\begin{figure*} 
  \resizebox{1.\textwidth}{!}{%
    \includegraphics{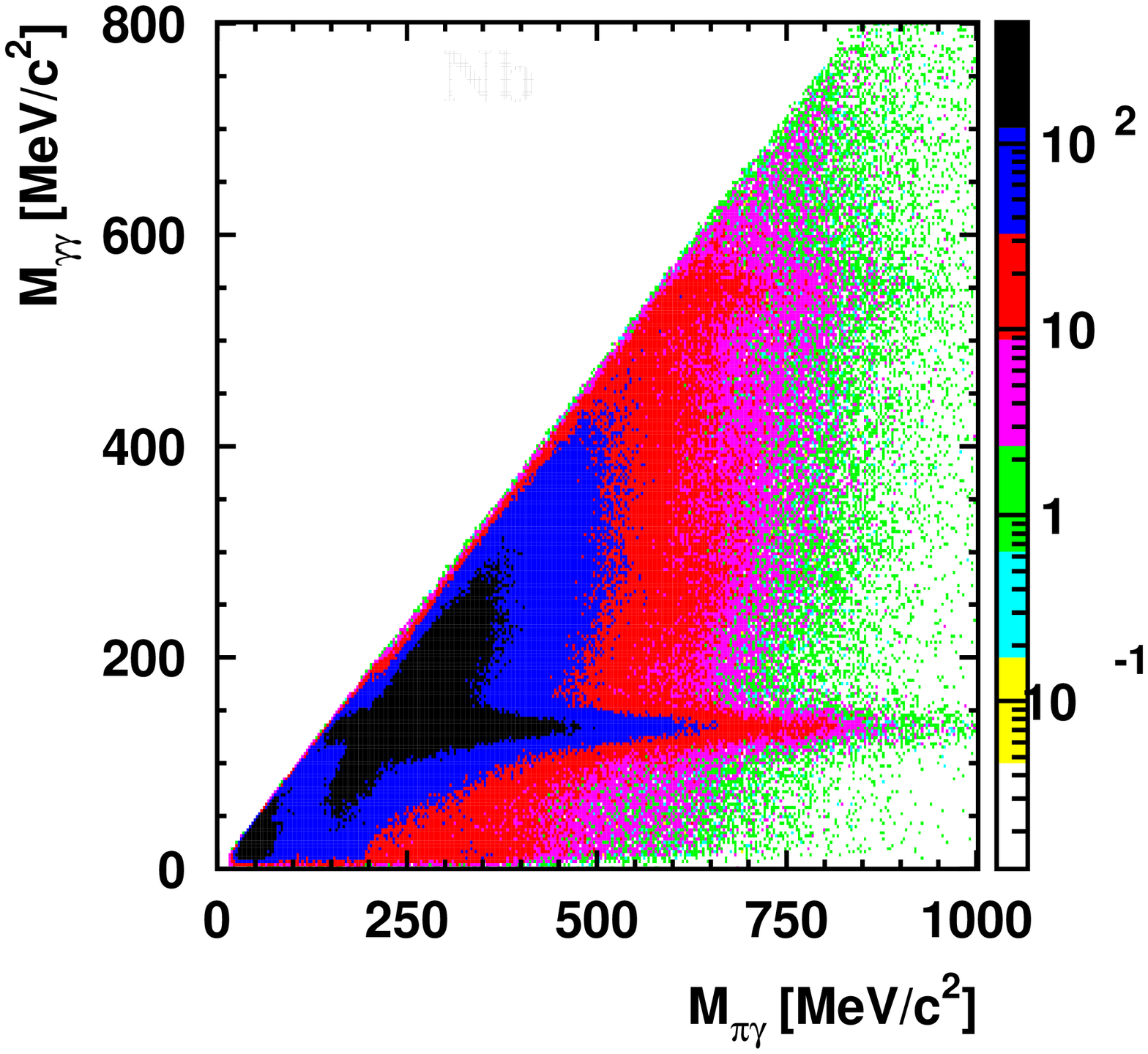}\includegraphics{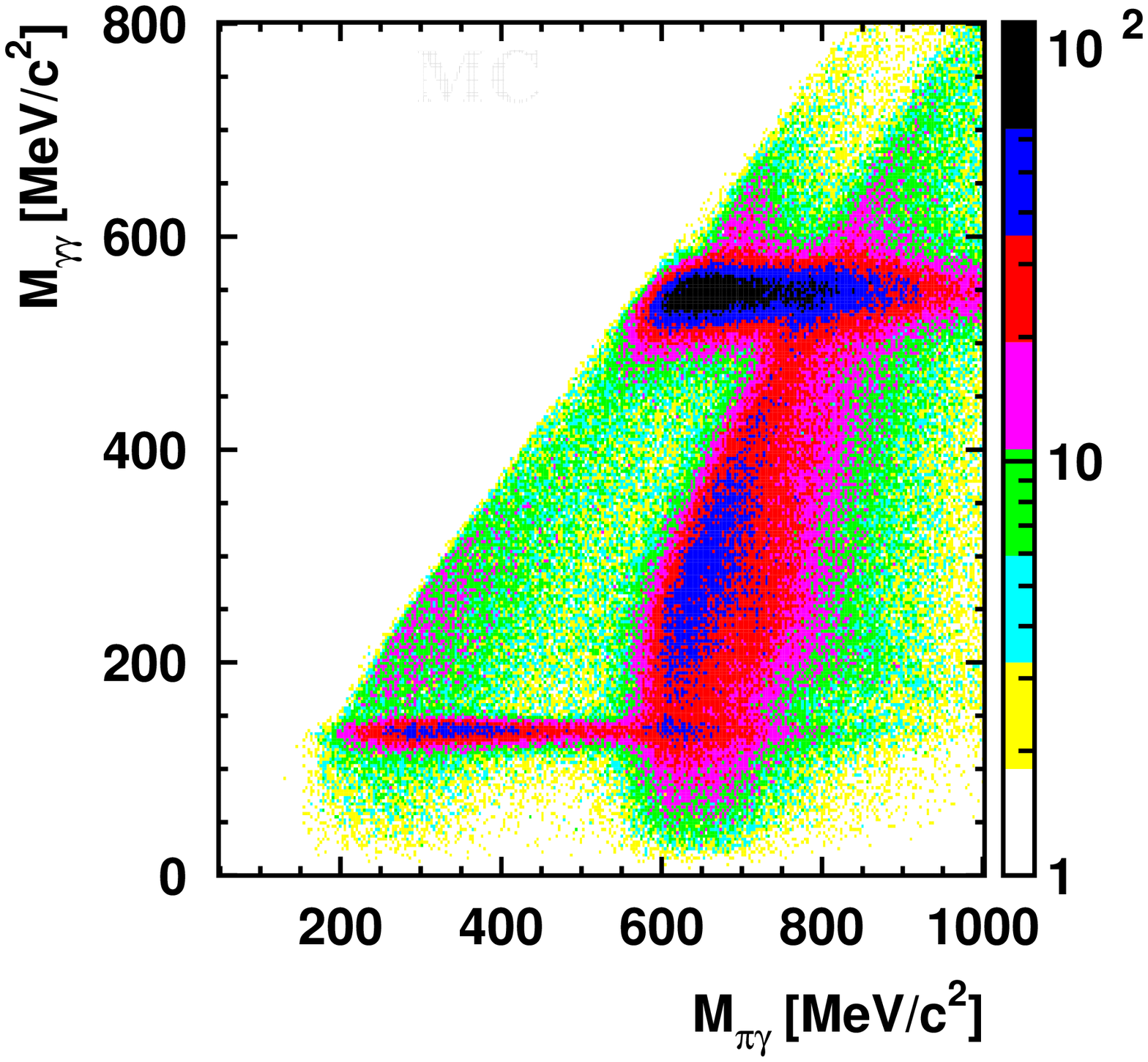}}
\caption{(Color online) Left: Experimental data for the reaction $\gamma Nb \rightarrow X p 3 \gamma$: the invariant mass of 2 photons (all 3 $\gamma\gamma$ combinations)
versus the $\pi^0 \gamma$ invariant mass. On the x-axis only one value is plotted per event for the 3$\gamma$ combination with the best $\pi^0$.  Right: Monte Carlo simulation: corresponding plot for the reaction $\gamma p \rightarrow p \pi^0 \eta$. Only those events are plotted where 1 proton and 3 photons are registered.} \label{fig:mgg_mggg}
\end{figure*} 

\subsubsection{\label{sec:level10}Time coincidence}

For reconstructing the reaction $\gamma A \rightarrow (A-1) p \omega
\rightarrow (A-1) p \gamma \gamma \gamma$  a prompt coincidence between a
particle in TAPS and an electron in the tagger was required to eliminate time
accidental background. Random time coincidences were subtracted using events
outside the prompt time coincidence window. For this analysis the prompt peak
was between -3 and 7 ns (Fig.~\ref{fig:time}a) and only events within the
prompt peak were accepted as candidates for the reactions of interest. The
asymmetric time cut allowed photons as well as nucleons to trigger the
event. Photons registered in TAPS were required to be prompt within -2.5 to
2.5 ns (Fig~\ref{fig:time}b). 

\subsubsection{\label{sec:level11}Split-off}
Monte Carlo simulations have shown that there is a strong contribution to the
3$\gamma$ invariant mass spectrum from single $\eta$ photoproduction $\eta
\rightarrow \gamma \gamma$, which has a huge cross section
at energies 900-1100 MeV. Shower fluctuations may result in an additional isolated energy deposit which is then reconstructed as an additional photon.  Due to this split-off of one photon cluster, 3 PED's are registered. With a high probability this process occurs in the transition
zone between the detectors TAPS and Crystal Barrel (Fig.~\ref{fig:exp} left). The photons from
split-off events are in most cases of low energy. It is possible to suppress
such events by applying the following cuts:\\ 
 1) The detectors in the TAPS to CB transition zone at angles
 between 26$^o$ and 34$^o$ are excluded from the analysis, as well as the detectors
 for $\theta >$ 155$^o$. According to simulations this leads to a 15$\%$ loss in the $\omega$ signal.\\
 2) The energy threshold in each photon cluster is set to 50 MeV.\\ 
As a result of both cuts the background is reduced by 21\%.\\

\subsubsection{\label{sec:level12}$\omega$ reconstruction}
The $\omega$ meson was reconstructed and identified via the three photon final
state invariant mass. Since the $\omega$ meson sequentially decays according to $\omega
\rightarrow \pi^{0} \gamma \rightarrow \gamma \gamma \gamma$, the
reconstructed particle can only be an $\omega$ meson if two of the three
photons stem from a $\pi^{0}$ decay. According to the relation 
\begin{equation}
E^{2} = m^{2} + p^{2}
\end{equation}

the $\pi^0$ and $\omega$ mesons have been identified via

\begin{equation}
m_{\pi^0} = \sqrt{2 E_{\gamma 1} E_{\gamma 2} (1- cos \theta)};
\end{equation}
\begin{equation}
m_{\omega} = \sqrt{(E_{\pi^0} + E_{\gamma 3})^{2} -
  (\overrightarrow{p}_{\pi^0} +\overrightarrow{p}_{\gamma 3} )^{2}} 
 \end{equation}
 
Thus in a two-dimensional plot of
the two photon invariant mass (all 3 combinations) against the $\pi^0 \gamma$
invariant mass, the $\omega$ meson should appear in this plane at the $\pi^{0}$
mass (2$\gamma$ axis) and the $\omega$ mass ($\pi^0 \gamma$ axis). Such a plot is
shown in Fig.~\ref{fig:mgg_mggg} left, where all cuts described so far have already been applied. 
\begin{figure*} 
  \resizebox{0.8\textwidth}{!}{%
    \includegraphics{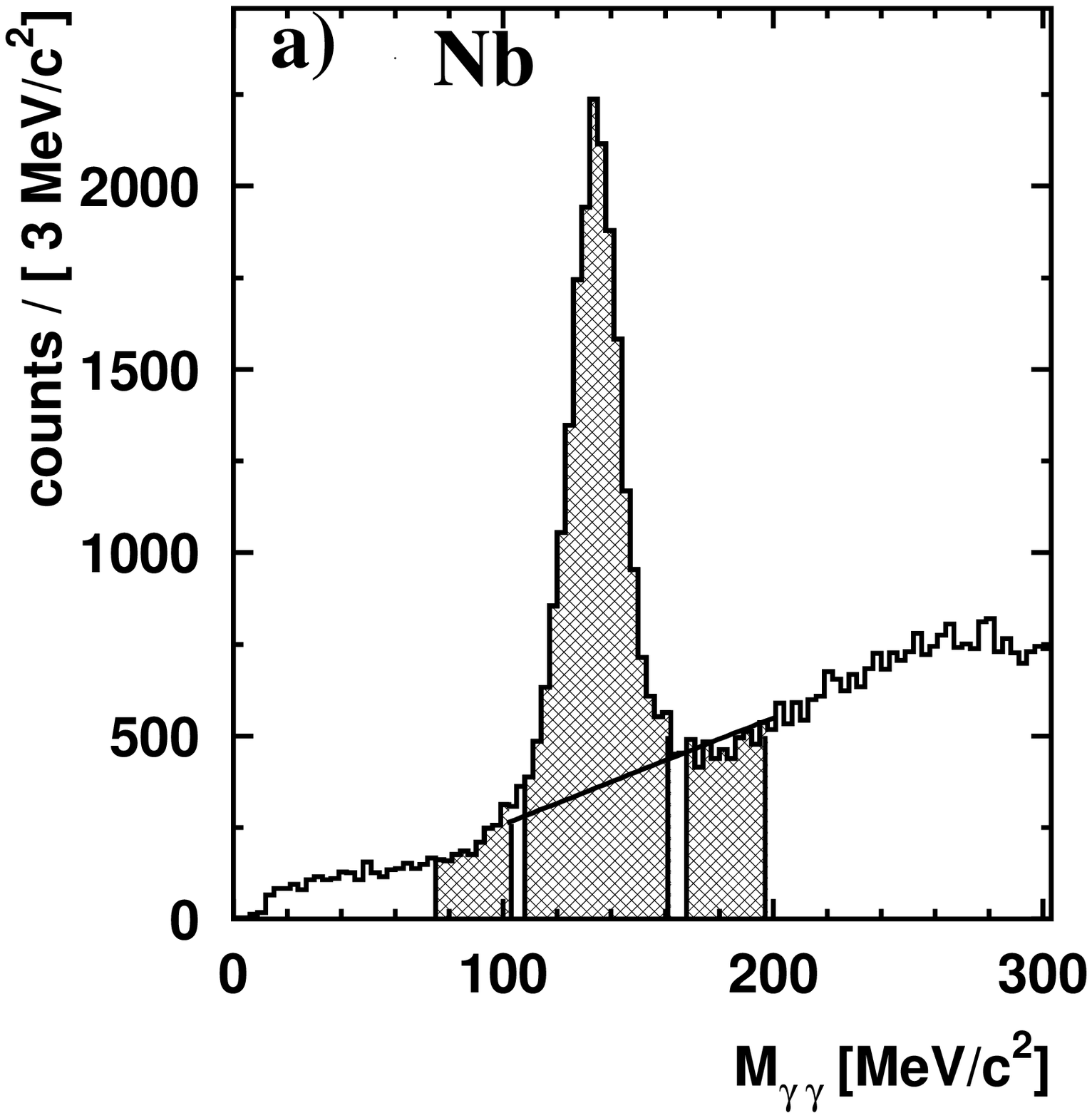}\includegraphics{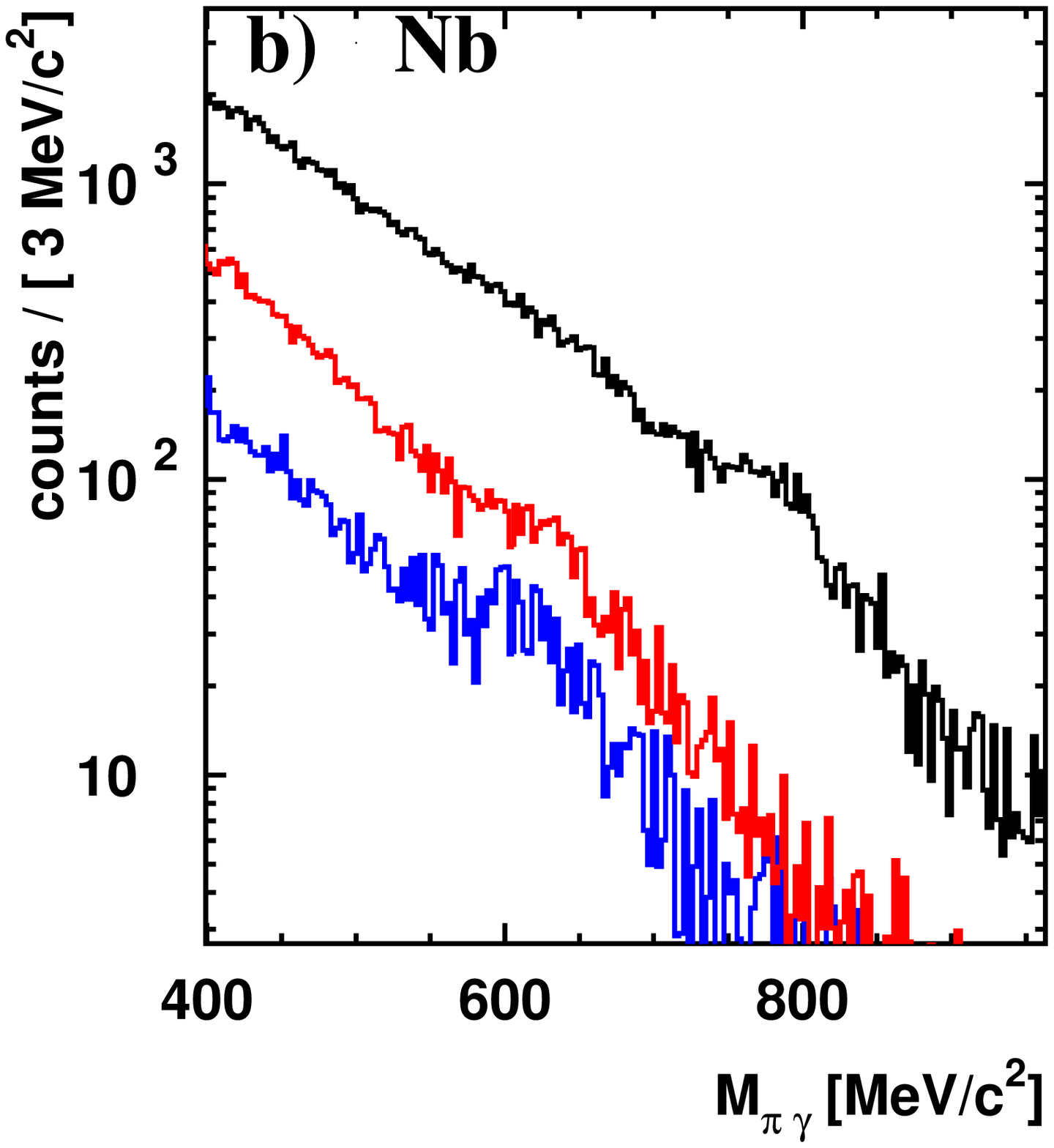}}     
  \resizebox{0.8\textwidth}{!}{%
    \includegraphics{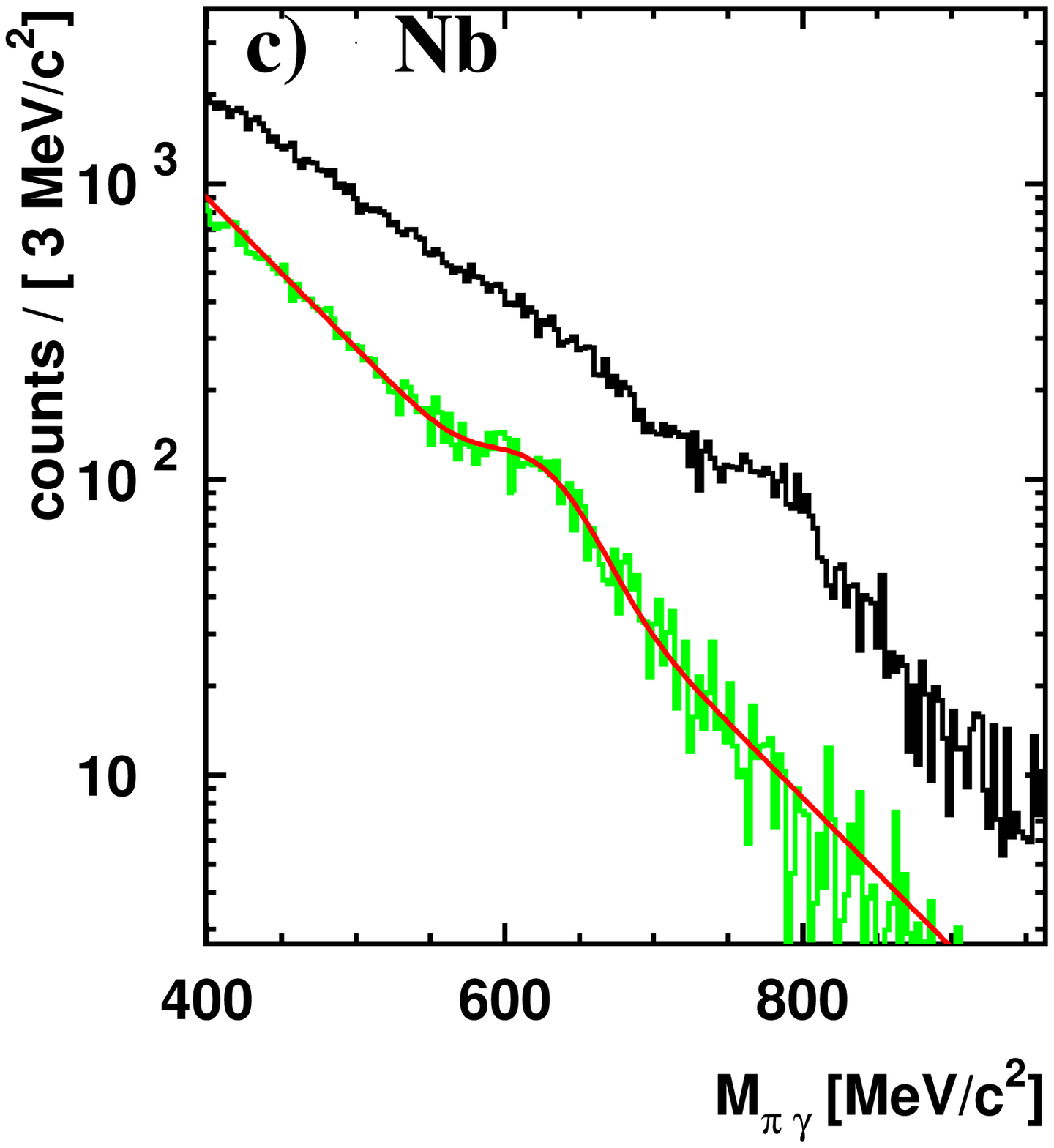}\includegraphics{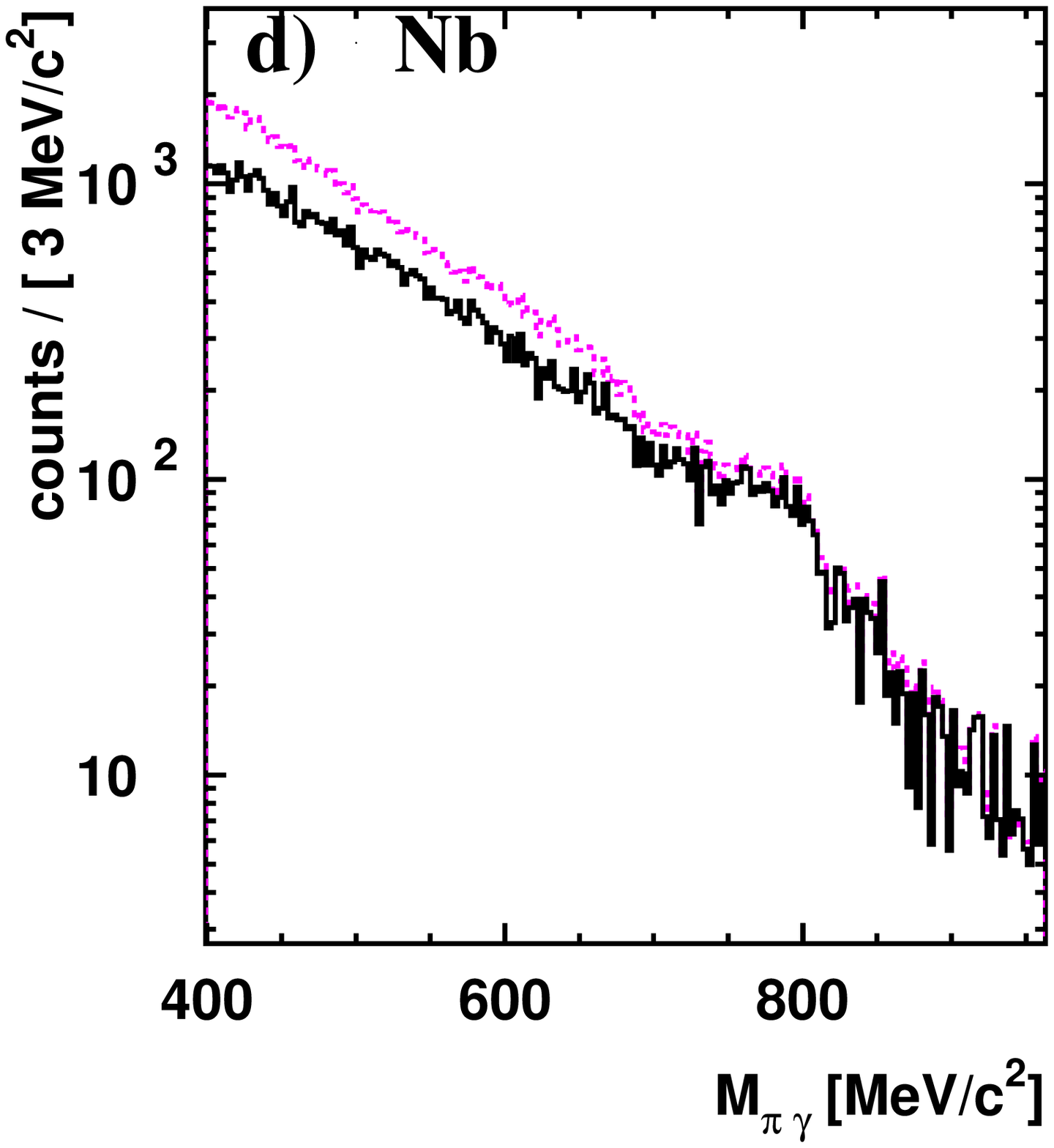}}
\caption{(Color online) a) Invariant mass of two $\gamma$'s; y-projection of Fig.~\ref{fig:mgg_mggg} left for a
  cut of M($\pi^{0} \gamma$) between 570 and 630 MeV. The shaded areas show
  the cuts for sideband subtraction. b) The M($\pi^{0} \gamma$) invariant mass
  distribution for the $\pi^{0}$ peak (black) and left (blue) and right(red)
  from the peak position as shown in a). c) The $\pi^{0} \gamma$ invariant
  mass in the $\pi^{0}$ peak (black) and the sum of the M($\pi^{0} \gamma$)
  projections left and right from the peak. The solid curve is a fit to the
  summed background spectrum. d) The $\pi^{0} \gamma$ invariant mass
  distribution after side band subtraction (solid histogram) compared to the spectrum without sideband subtraction (dashed histogram). All spectra refer to the $Nb$ target.}\label{fig:sideband} 
\end{figure*}

\subsubsection{\label{sec:level13}Sideband subtraction technique}
As mentioned above, one of the channels which contribute to the background is
\begin{figure*}
\resizebox{1.0\textwidth}{!}{
\includegraphics{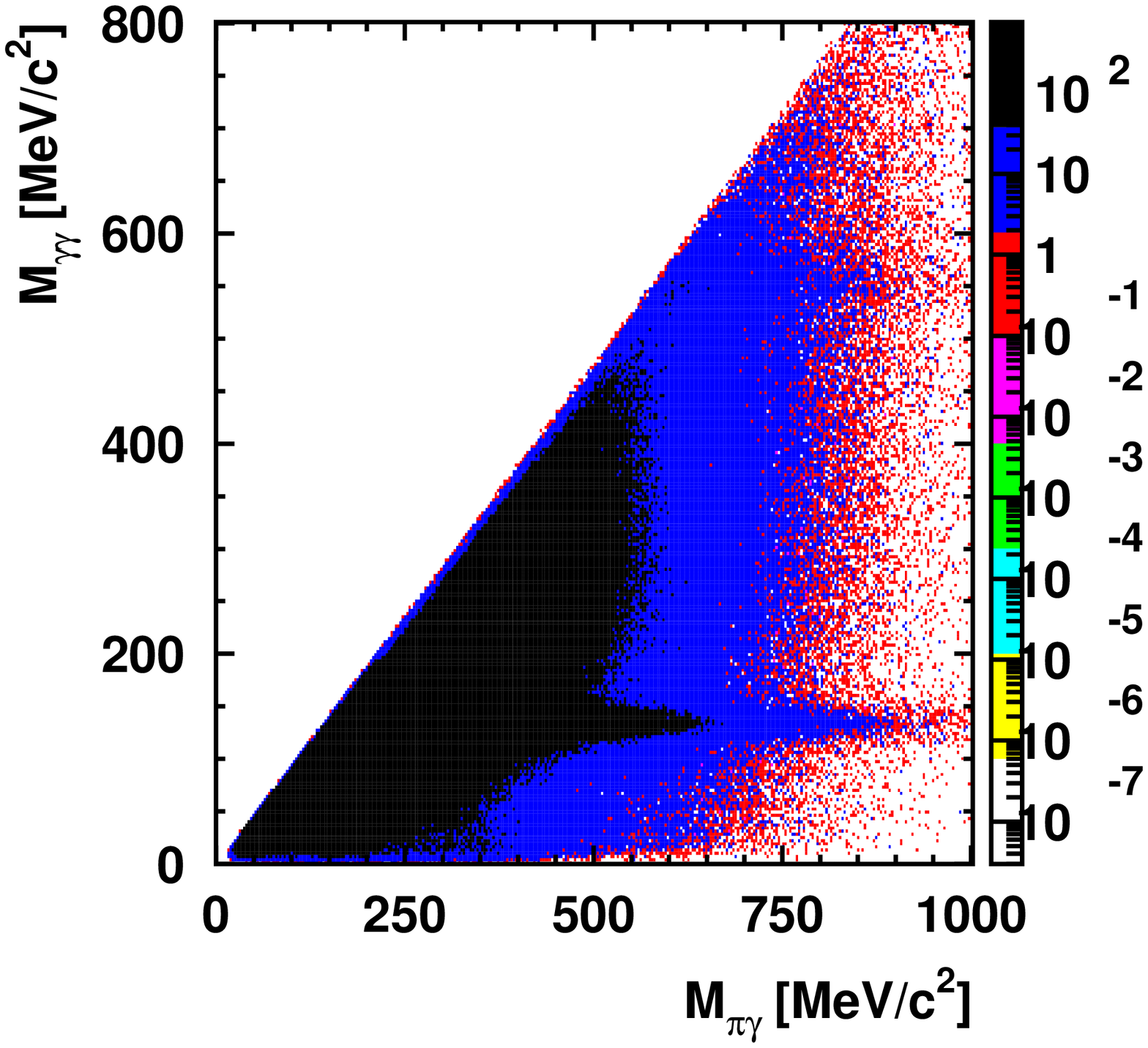}\includegraphics{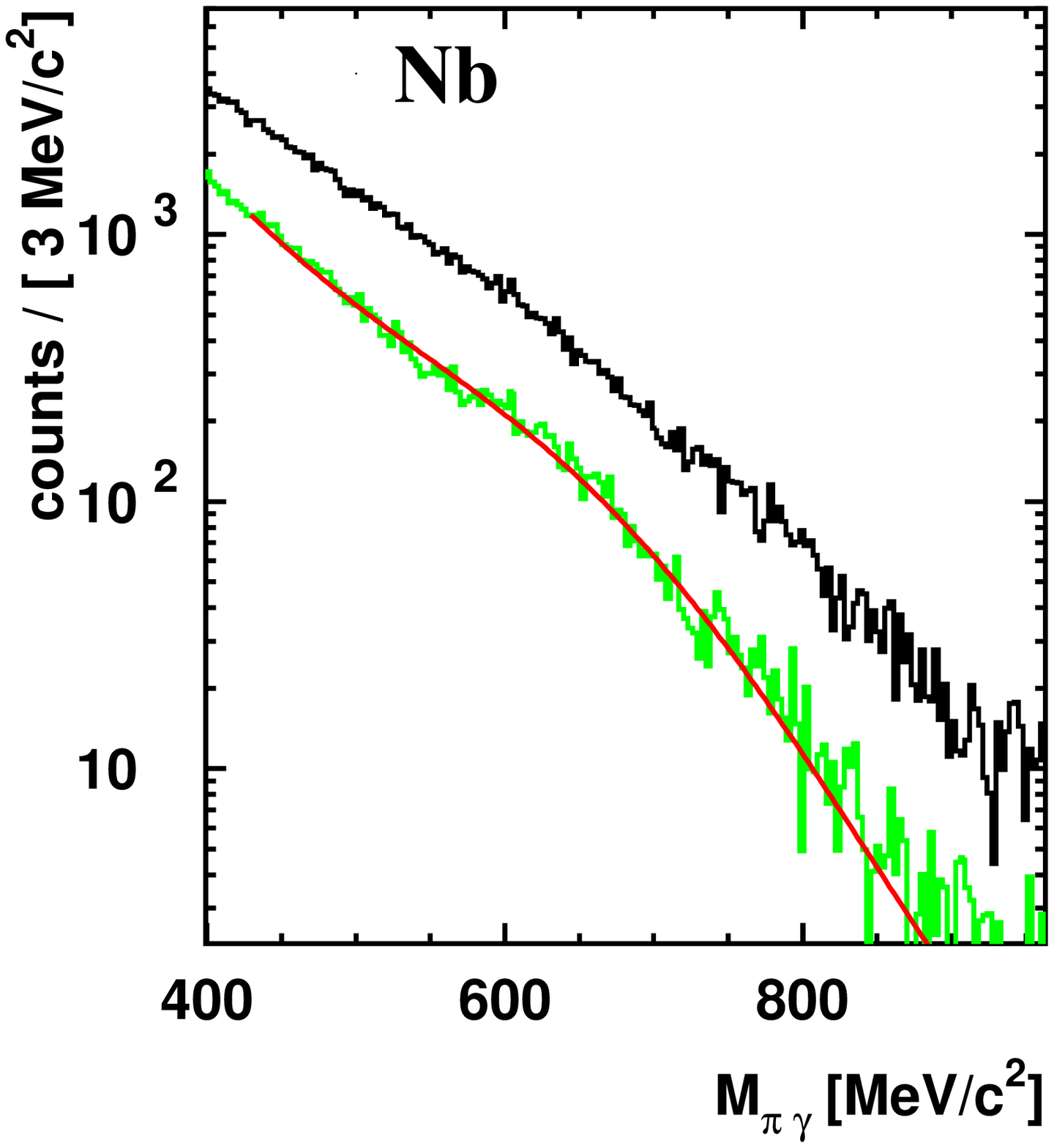}}
\caption{(Color online) Left: Experimental data for the reaction $\gamma Nb \rightarrow X p 4 \gamma$: after omitting one of the photons the invariant mass of 2 photons is plotted
versus the $\pi^0 \gamma$ invariant mass. On the x-axis only one value is plotted per event for the 3$\gamma$ combination with the best $\pi^0$. Right: The $\pi^{0} \gamma$ invariant
  mass in the $\pi^{0}$ peak (black) and the sum of the M($\pi^{0} \gamma$)
  projections left and right from the peak. The solid curve is a fit to the
  summed background spectrum. }
\label{fig:bg_ref}
\end{figure*}
$\pi^{0} \eta$ photoproduction with 4 photons in the final state when one of
the photons escapes detection. In order to suppress this background in the
$\pi^0 \gamma$ spectrum the technique of side band subtraction was used.
Monte Carlo simulations (Fig.~\ref{fig:mgg_mggg} right) show that events obtained by combining 2
$\gamma$'s from an $\eta$ decay with a $\gamma$ coming from a $\pi^0$ decay appear as an
almost vertical band around 600 MeV on the x-axis, which shows up as
a bump in the M($\pi^{0} \gamma$) projection.
\begin{figure*} 
  \resizebox{0.8\textwidth}{!}{
    \includegraphics{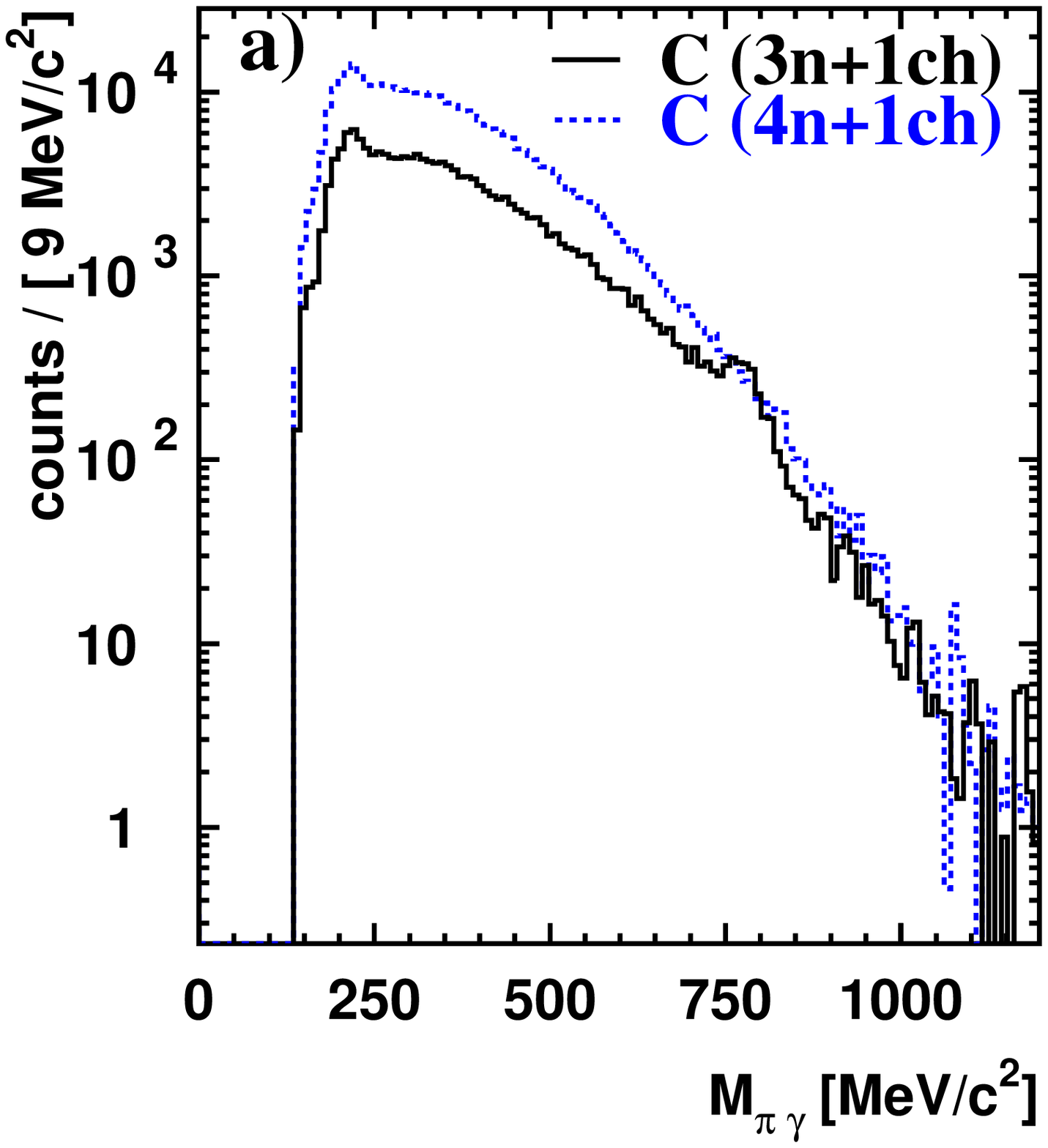}\includegraphics{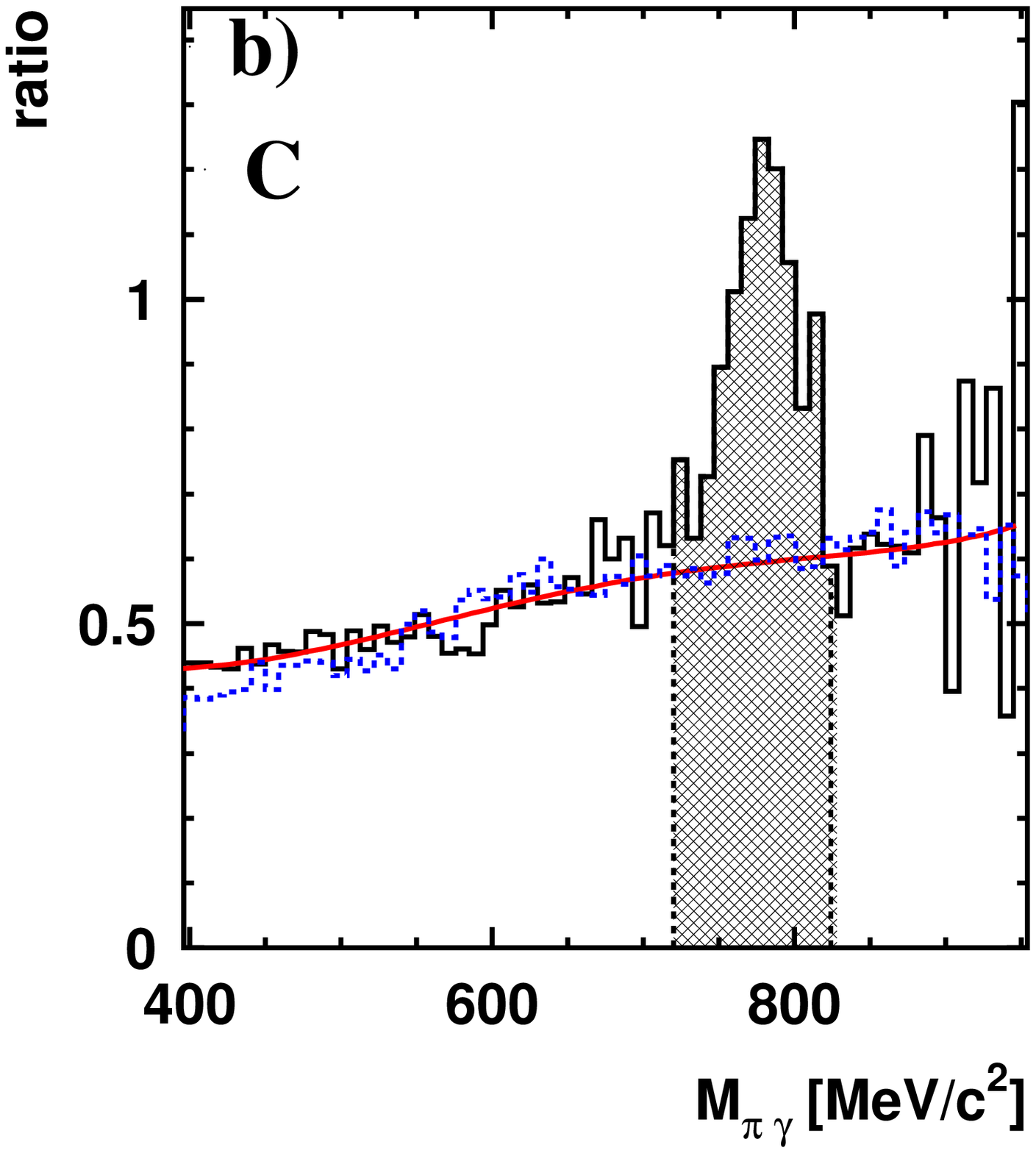}}  
  \resizebox{0.8\textwidth}{!}{
     \includegraphics{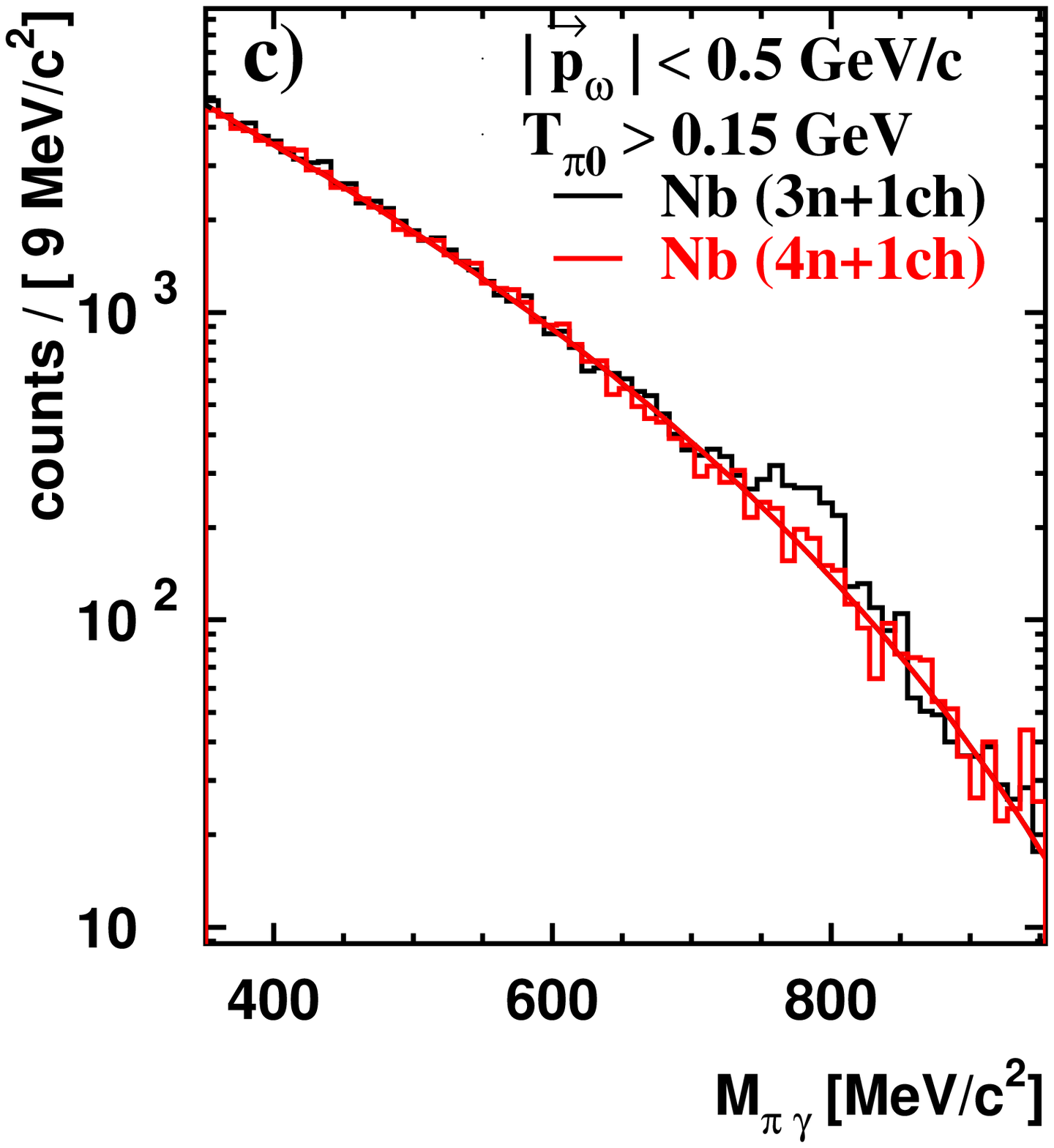}\includegraphics{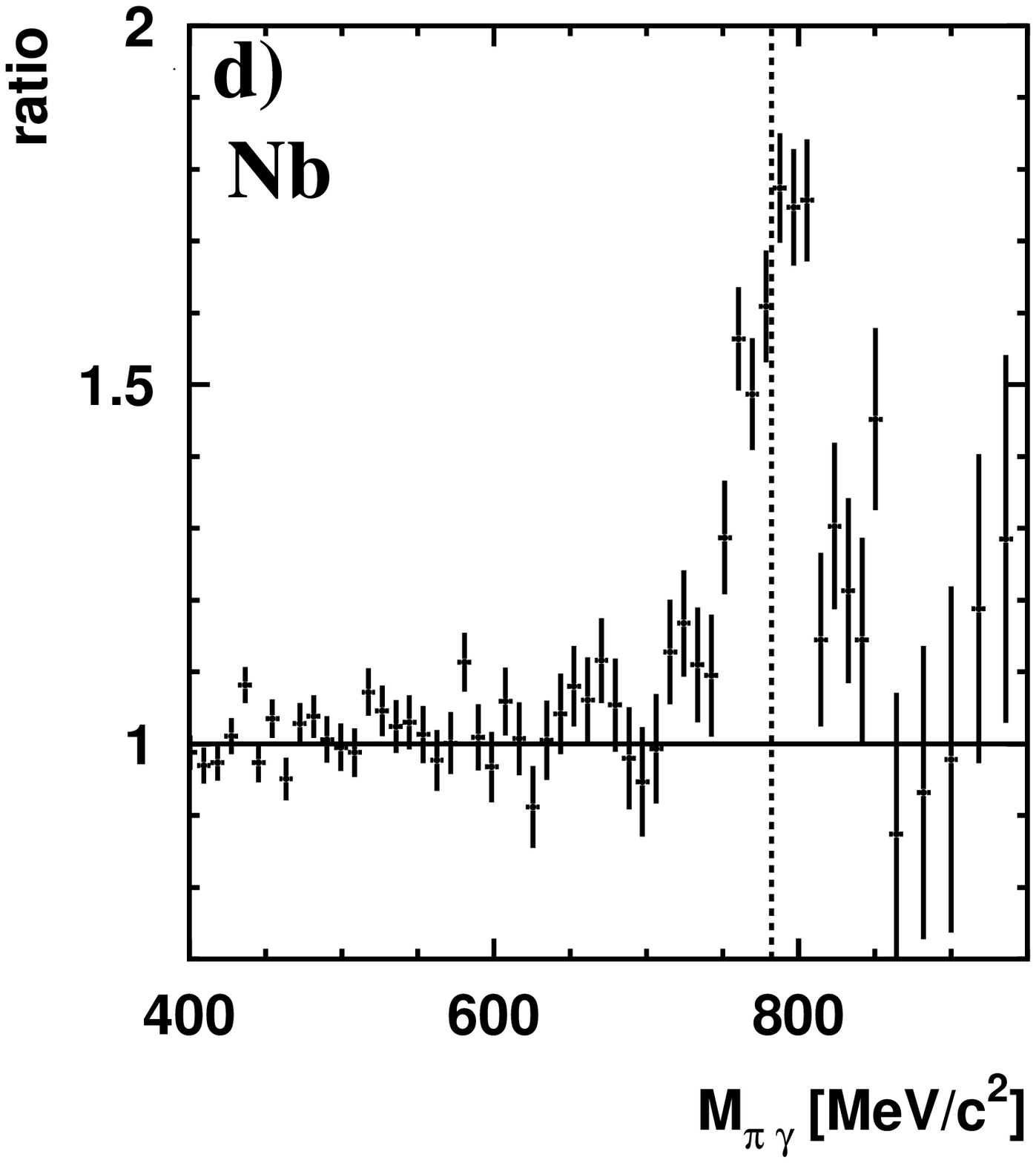}} 
\caption{(Color online) a) $\pi^0\gamma $ signal (solid curve) and background spectrum  (dotted curve) for the $C$ target deduced from events with 4 neutral and 1 charged hit. b)
    Correction function derived from the carbon data. The dashed curve shows the mass
    dependence of this correction expected from a simulation of the 2$\pi^0$
    channel. c) The $\pi^{0} \gamma$ signal spectrum and the corrected and
    normalized background spectrum for the $Nb$ target. The solid curve
    represents a fit to the background distribution. d) Ratio of the $\pi^{0}
    \gamma$ spectrum to the background spectrum for the $Nb$ target generated
    from events with 4 neutral and 1 charged hit.}
    \label{fig:bg} 
\end{figure*} 
 
To reduce this bump and to suppress the combinatorial background, side band subtraction has been applied. Fig.~\ref{fig:sideband}a shows the projection on the y-axis
M($\gamma \gamma$) for the mass range 570 $\le$ M($\gamma \gamma \gamma$)
$\le$ 630 MeV. Projections on the x-axis M($\pi^{0} \gamma$) are shown in
Fig.~\ref{fig:sideband}b for cuts close to the pion mass: 110 to 160 MeV and
left (75 to 100 MeV) and right(170 to 195 MeV) from the peak. The sum of 
both sideband spectra (Fig.~\ref{fig:sideband}c) was normalized to the background counts
under the pion peak and fitted with an exponential and Gaussian function. In
the next step this curve was subtracted from the M($\pi^{0} \gamma$) spectrum
over the full mass range. Fig.~\ref{fig:sideband}d shows the resulting
spectrum after the sideband subtraction. The bump around 600 MeV is removed from the final spectrum. The background in the spectrum for masses of
400 MeV to 700 MeV is 37\% lower compared to the spectrum without sideband
subtraction, but the difference in the region of the $\omega$ signal from 700 MeV
to 820 MeV is only 14\% (Fig.~\ref{fig:sideband}d). It is essential to remove this structure arising from the $\pi^0 \eta$ channel as it extends towards higher masses where it may distort the $\omega$ line shape.\\

\subsubsection{\label{sec:level14}Momentum cut}
Only $\omega$ mesons decaying inside the nucleus carry information on the
in-medium properties which are to be studied. To enhance the in-medium decay
probability, the vector meson decay length should be comparable to nuclear
dimensions. This was achieved in the analysis by applying a kinematic cut on
the three momentum of the $\omega$ meson $\arrowvert
\overrightarrow{p_{\omega}}\arrowvert \leq$  500 MeV/c. But still, only a
fraction of the $\omega$ mesons will decay inside the nucleus. Thus, one
expects the $\pi^{0} \gamma$ invariant mass spectrum to show a superposition of
decays outside of the nucleus at the vacuum mass with a peak position at 782 MeV/$c^2$ and of possibly modified decays inside the nucleus~\cite{johan}.  In addition, the most pronounced in-medium effects are expected for low meson momenta with respect to the the nuclear medium.

\subsubsection{\label{sec:level15}{Cut on the kinetic energy of the $\pi^{0}$ in the final state}} 
The disadvantage of reconstructing the $\omega$ meson in the decay mode
$\omega \rightarrow \pi^{0} \gamma$ is a possible rescattering of the
$\pi^{0}$ meson which was studied in ~\cite{johan}. The authors have
demonstrated that the constraint on the pion kinetic energy $T_{\pi^0} >$ 150
MeV suppresses the final state interaction down to the percent level in the
invariant mass range of interest (650 MeV $\leq M(\pi^0 \gamma) \leq 850 $
MeV). This result has been confirmed in transport calculations ~\cite{Kaskulov},~\cite{Muhlich}.  

\begin{figure*} 
  \resizebox{1.0\textwidth}{!}{
    \includegraphics[height=1.0\textheight]{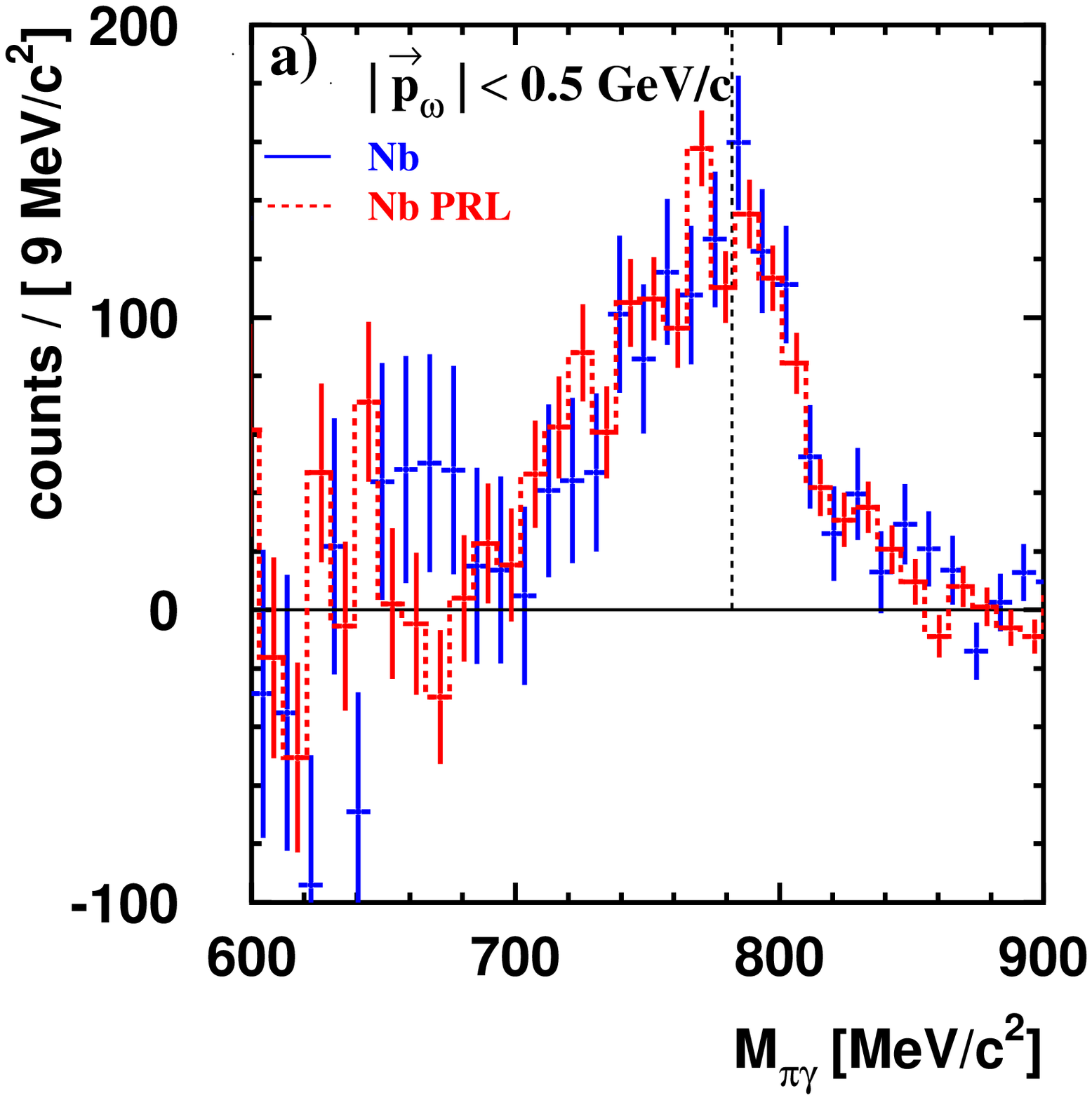}
    \includegraphics[height=1.0\textheight]{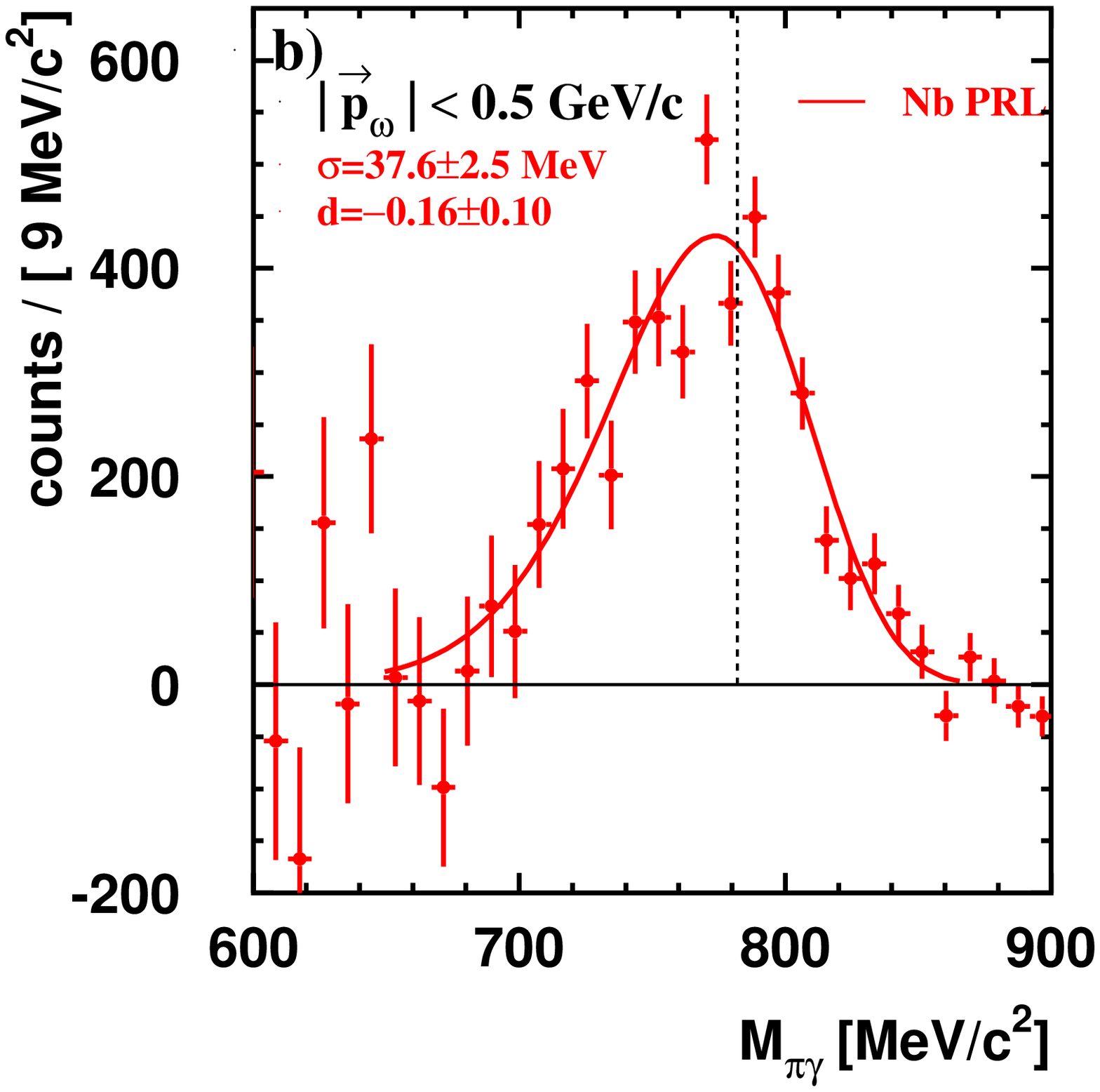}
    \includegraphics[height=1.0\textheight]{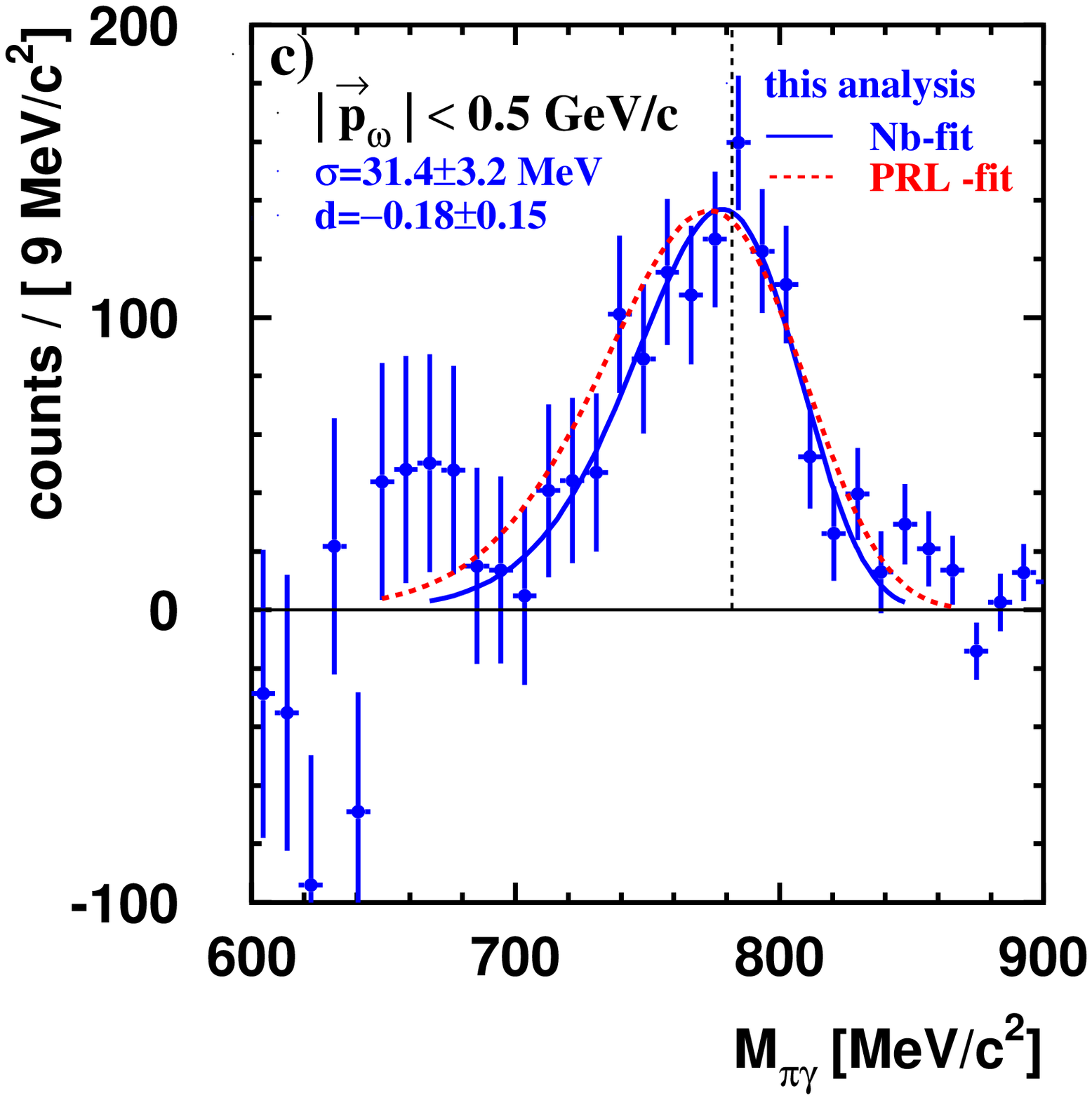}}
\caption{(Color online) a) $\omega$ signal for the $Nb$ target from this analysis (solid points)
    compared to the $\omega$ signal published in \cite{david} (dashed). For
    the comparison of the line shapes the latter data have been scaled down by
    factor of 3.3 to match the intensity of the signal in the current analysis
    where more restrictive cuts have been applied. The spectra are without cut on the pion kinetic    energy. b) fit to the $\omega$ signal published in \cite{david} using the function given by 
    Eq.~\ref{novis_1}. c) fit to the $\omega$ signal obtained in this analysis using the same function. The fit to the signal from \cite{david} is included for comparison as red dashed line.}
    \label{fig:omega_signal} 
\end{figure*}   

\subsection{\label{sec:level16}Background Analysis}
The next main step in the analysis was the determination of the background
directly from the data and its absolute normalization.  

\subsubsection{\label{sec:level17}{Background reconstruction}}
As mentioned before, the most probable sources of background come from the
reactions $\gamma A \rightarrow (A-1) p \pi^{0} \pi^{0}$ and $\gamma A
\rightarrow (A-1) p \pi^{0} \eta$ with 4 $\gamma$ and one proton in the final
state. Due to photon cluster overlap or detection inefficiencies one of the
four photons may not be registered, thereby giving rise to a $\pi^{0} \gamma$
final state, which is exactly identical and therefore not distinguishable from
the $\omega$ meson final state. To study this background, 5 PED events were
selected with 4 neutral and 1 charged hit. One of the four neutral particles
was randomly omitted and from the remaining photons a $\pi^{0}$ was identified
and combined with the 3rd photon. The 2-dimensional plot of mass
$M_{\gamma\gamma}$ versus the $\pi^{0} \gamma$ invariant mass is shown in Fig.~\ref{fig:bg_ref} (left). It is similar to
the plot from the 4 PED events for the $\omega$ reconstruction (see Fig.~\ref{fig:mgg_mggg} left). This is filled four times for all combinations with 4 photons. The side band subtraction technique was applied as described in sec. ~\ref{sec:level13}. The applied cuts on the  $\pi^{0} \gamma$ momentum, on the kinetic energy of the pion and on the prompt peak were the same as for the
$\omega$ meson reconstruction The resulting $\pi^{0} \gamma$ spectrum is shown in  
Fig.~\ref{fig:bg_ref} (right).  
\subsubsection{\label{sec:level18}{Lost photons}}
The slopes in the signal and background (BG) spectra shown in Fig.~\ref{fig:bg}a are different due to the different kinematics in detecting events with 4 neutral and 1 charged particle
with respect to events with 3 neutral and 1 charged hits, reflecting the energy dependence of the probability that only 3 out of 4 photons are detected. The ratio of both spectra is shown in Fig.~\ref{fig:bg}b for the $C$ target. A procedure has been developed to correct the background
slope in the $Nb$ spectrum using the data obtained on the carbon target which
is such a light nucleus that strong in-medium effects are not expected. The
correction function is derived by fitting the ratio of the spectra for the
carbon data excluding the peak region, as it is shown in Fig.~\ref{fig:bg}b. The dependence of this correction on the $\pi^0 \gamma$ invariant mass is
confirmed by simulations (dashed curve in Fig.~\ref{fig:bg}b) studying the
energy dependence of the probability to register only 3 out of 4 photons for the
dominating 2$\pi^0$ background channel. The $\pi^0 \gamma$ background for $Nb$
from events with 4 neutral and 1 charged particles is multiplied with this
correction function. As a result, the background for the $Nb$ data changes its
slope. 

\subsubsection{\label{sec:level19}{Background normalization}}

The absolute height of the background is determined by requesting the
same number of counts for the signal and background spectra in the mass range
from 400 to 960 MeV, excluding the counts in the $\omega$ peak which account
for only 2\% of the total yield in the given mass range. Thereby, the
background level is fixed without paying any attention to the $\omega$ signal
region. Fig.~\ref{fig:bg}c shows the $\pi^{0} \gamma$  and the corrected and
normalized background spectra. The ratio of these two spectra given in
Fig.~\ref{fig:bg}d demonstrates that the background in the $\pi^0 \gamma$
spectrum on $Nb$ is properly reproduced by the background spectrum generated
from the events with 4 neutral and 1 charged hits after applying the required
corrections. In the invariant mass range from 400 to 700 MeV the average
deviation from unity is 4$\%$. For higher invariant masses fluctuations become stronger
because of the poorer statistics.

\begin{figure*} 
  \resizebox{1.0\textwidth}{!}{
    \includegraphics{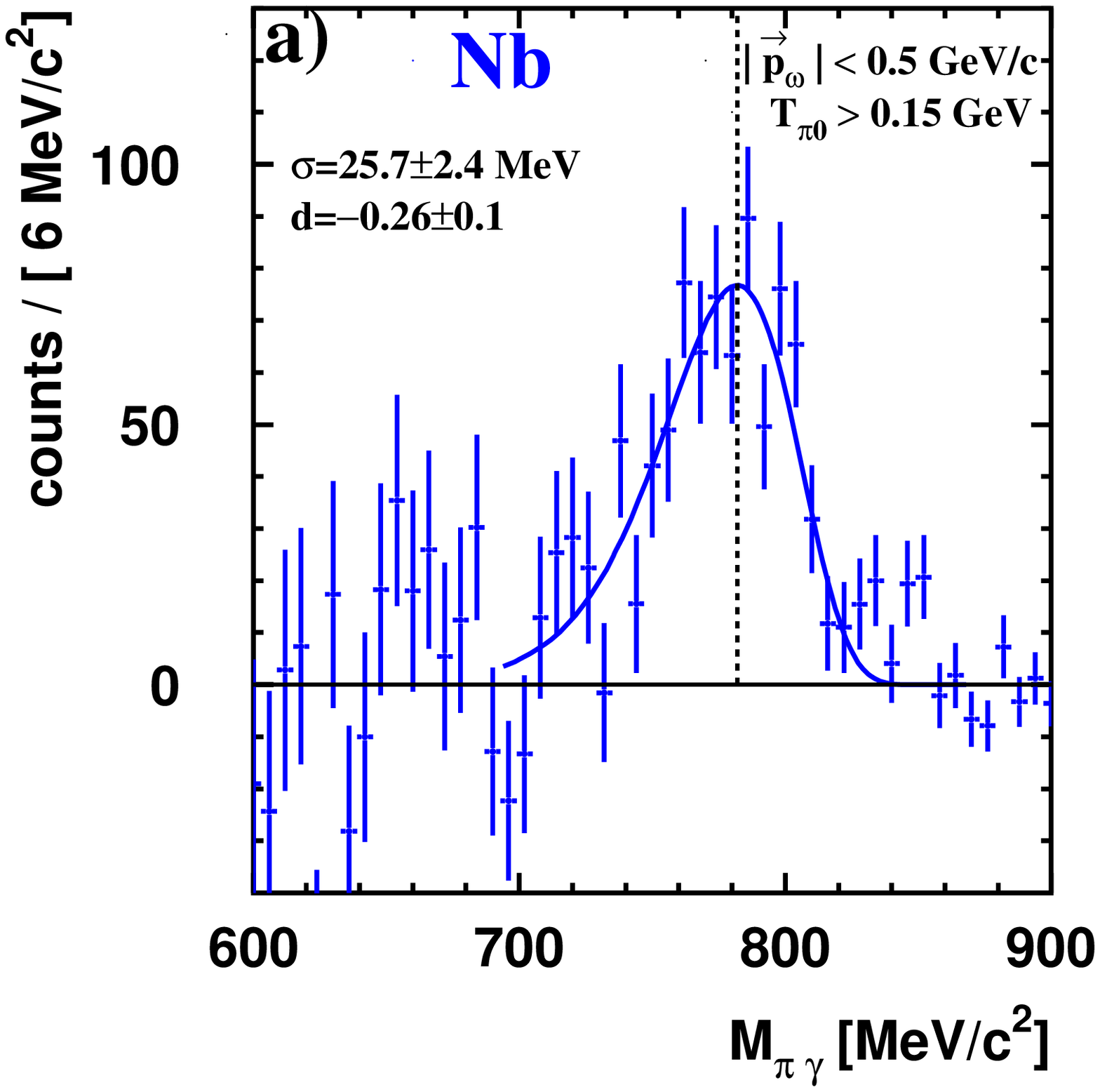}
    \includegraphics{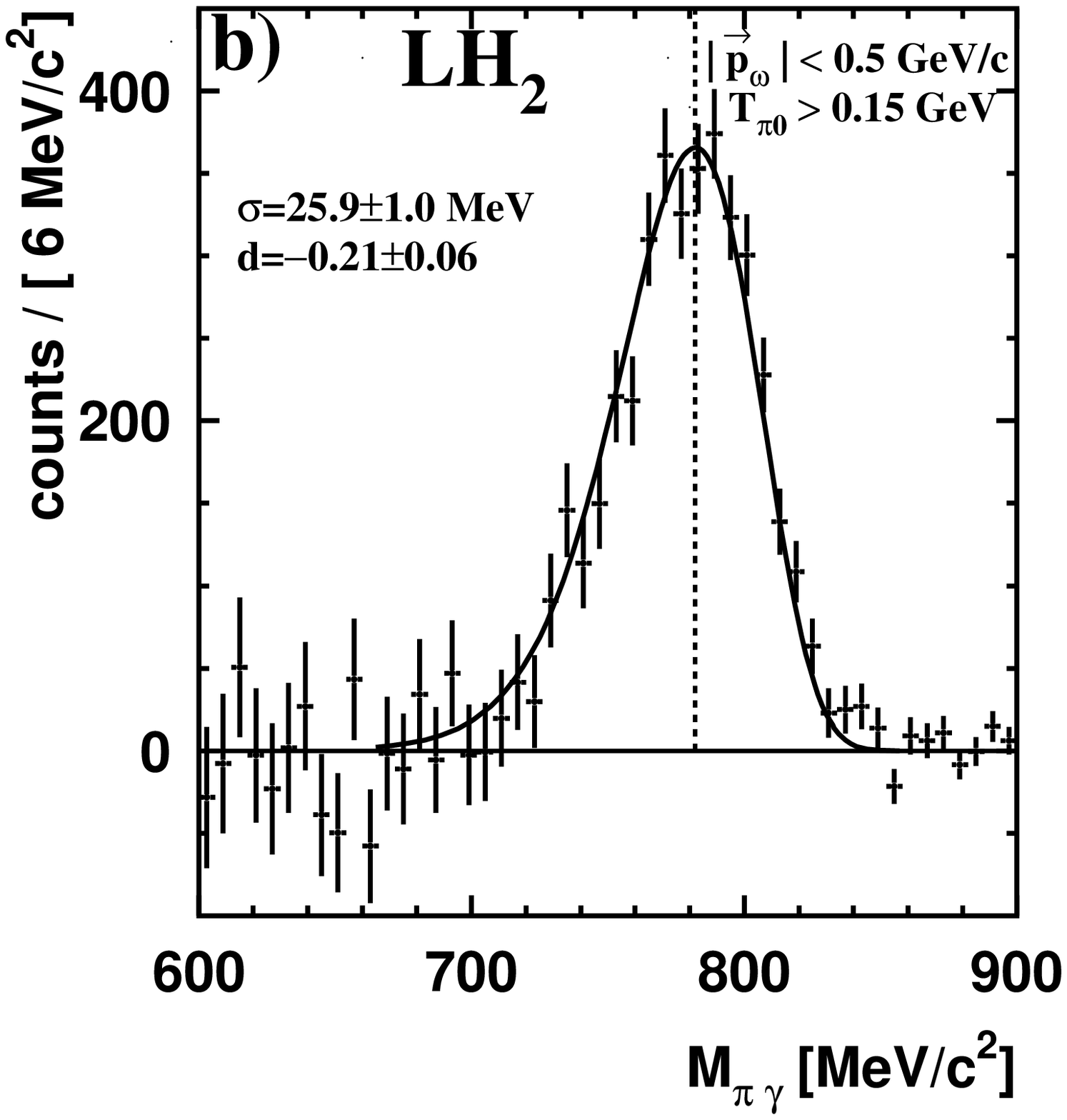}
    \includegraphics{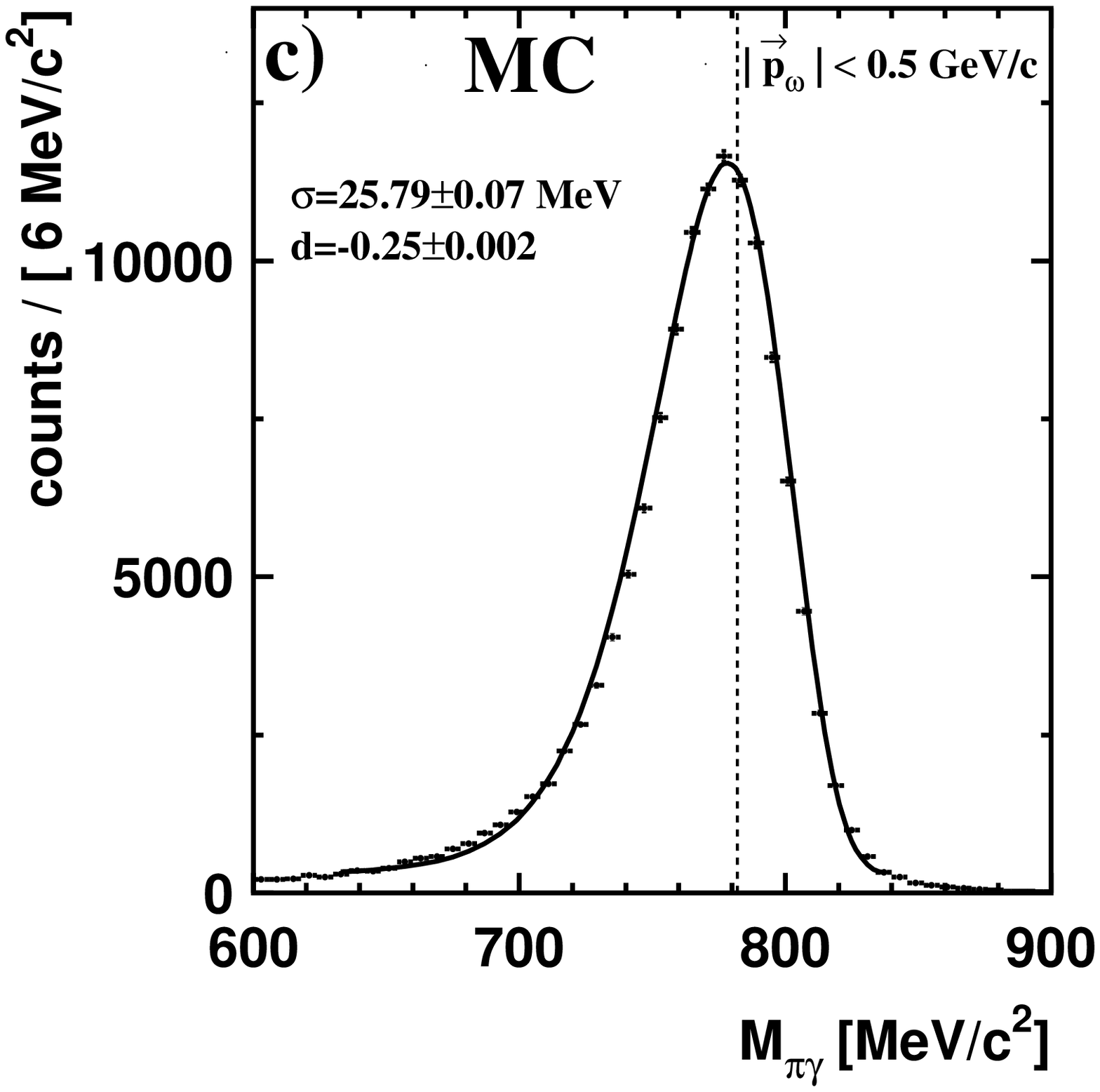}} 
\caption{(Color online) a) $\omega$ signal for $\pi^{0} \gamma$ momenta below 500 MeV/c and
    kinetic energy $T_{\pi^0} >$ 150 MeV ($Nb$ target). The solid curve
    represents a fit with the function of Eq.~\ref{novis_1}. b) $\omega$ signal for a $LH_{2}$ target and c)  $\omega$ signal from MC simulation.} \label{fig:signal_fits}
\end{figure*}  

\subsection{\label{sec:level20}Results and Discussion}

The $\omega$ signal shown in Fig.~\ref{fig:omega_signal}a is obtained by
subtraction of the background from the signal spectrum. For comparison the
$\omega$ line shape deduced in the previous analysis \cite{david} is overlayed.
Only slight differences are observed which, however, become more apparent
when the signals are fitted individually, as shown in Fig.~\ref{fig:omega_signal}b,c. The following function \cite{Aubert} has been 
used for the fits:
\begin{equation}
f(x) = A \cdot \exp\left[{-\frac{1}{2} \left(\frac{\ln{q_{x}}}{d}\right)^2 + d^{2}}\right] \label{novis_1}
\end{equation}
where
\begin{equation}
q_{x} = 1+ \frac{(x-E_p)}{\sigma} \cdot \frac{\sinh(d\sqrt{\ln{4}})}{\sqrt{\ln{4}}} \label{novis_2}
\end{equation}

Here $A$ is the amplitude of the signal, $E_{p}$ is the peak energy, $\sigma$ is
FWHM/2.35 and $d$ is the asymmetry parameter. This function takes into account
the tail in the region of lower invariant masses resulting from the energy
response of the calorimeters. Fig.~\ref{fig:omega_signal} compares the fit
to the $\omega$ signal published in \cite{david} (Fig.~\ref{fig:omega_signal}b) with the fit to the 
$\omega$ signal obtained in this work (Fig.~\ref{fig:omega_signal}c). In the re-analysis a somewhat narrower $\omega$ signal is observed.

Applying in addition the cut on the kinetic energy of the $\pi^0$ meson ($T_{\pi^0} > $150 MeV)
a fit to the $\omega$ signal (Fig.~\ref{fig:signal_fits}a) yields a width parameter
$\sigma$=25.7$ \pm$ 2.4 MeV which is consistent within
errors with the $LH_2$ and MC signals (Fig.~\ref{fig:signal_fits}b,c) which serve as a reference. The deviation from the reference signals claimed in \cite{david} and interpreted as evidence
for an in-medium mass shift of the $\omega$ meson is thus not confirmed in the
re-analyis of the data described in this paper. The current analysis does not
yield any evidence for an in-medium lowering of the $\omega$ mass. This does
not necessarily mean that there is no mass shift because the $\omega$ line
shape may be insensitive to in-medium modifications as pointed out in ~\cite{Kaskulov}. \\

\begin{figure} 
  \resizebox{0.5\textwidth}{!}{
    \includegraphics{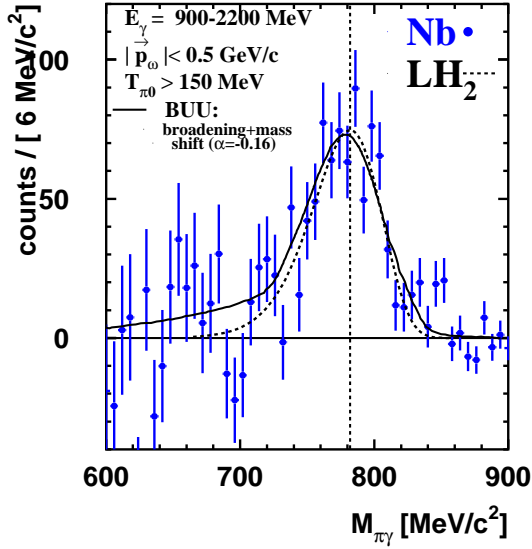}}  
\caption{(Color online) $\omega$ signal for the $Nb$ target from this analysis (solid points)
    in comparison to the $\omega$ line shape measured on a $LH_2$ target (dashed curve) and to
    a GiBUU simulation \cite{Muehlich_PhD} (solid curve) assuming a mass shift by -16$\%$ at normal nuclear
    matter density.} \label{fig:GiBUU_comp}
\end{figure}

This problem is illustrated in Fig.~\ref{fig:GiBUU_comp} which 
compares the $\omega$ line shape of the present analysis to the line shape for
the $LH_2$ target as well as to a prediction of the $\omega$ line shape in a
GiBUU transport model calculation. In this calculation an in-medium pole mass
shift according to  
\begin{equation}
 m_{\omega}^{*} = m_{\omega}^{0}(1-0.16 \frac{\rho_{N}}{\rho_{0}})
\end{equation}
has been assumed. Here, $\rho_N$ is the nuclear density at the decay point of
the $\omega$ meson and $\rho_0$ is the normal nuclear matter density. The fact
that the experimental signal is consistent with both scenarios indicates that
the line shape is indeed insensitive to in-medium modifications for the
given invariant mass resolution and statistics.\\
  This insensitivity is first of all due to the relatively long lifetime of the $\omega$ meson. Even requiring the $\omega$ recoil momentum to be lower than 500 MeV/c, only about 20\% of all $\omega \rightarrow \pi^0 \gamma$ decays in $Nb$ occur at densities $\rho/\rho_0 > 0.1$ for the given reaction kinematics according to BUU simulations \cite{Muehlich_PhD}. In addition, due to inelastic processes like $\omega N \rightarrow \pi N$, the $\omega$ mesons are removed in the nuclear medium thereby reducing their effective lifetime and correspondingly increasing their width. If this broadening is very large as observed for  the $\omega$ meson \cite{Kotulla} the strength of the in-medium signal is spread out in mass so strongly that it becomes hard to distinguish it from the background. \\
  
  This argument can also be formulated more rigorously as discussed in \cite{Lehr,LMM}. Any mass distribution measurement of a vector meson $V$ from its decay into particles $p_1, p_2$ does not give the hadronic spectral function of the meson directly but folded with the branching ratio $\Gamma_{V \to  p_1 + p_2}/\Gamma_{\rm tot}$ into the specific final channel one is investigating ~\cite{eichstaedt}
\begin{equation}
\frac{d\sigma_{\gamma N \to N(p_1,p_2)}}{d\mu} = \frac{d\sigma_{\gamma N \to VN}}{d\mu} \times \frac{\Gamma_{V \to  p_1 + p_2}}{\Gamma_{\rm tot}}(\mu).\label{dsigmamufinal}
\end{equation}
Since the branching ratio may depend on the mass $\mu$ the unfolding is not
trivial. Integrating Eq.~\ref{dsigmamufinal} over all nucleons and parameterizing the spectral function $A(\mu)$ by a Breit-Wigner function, the result involves the term    
\begin{equation}
A(\mu)\, \frac{\Gamma_{V \to \rm final\;state}}{\Gamma_{\rm tot}}
 = \frac{1}{\pi}\frac{\mu \, \Gamma_{\rm tot}}{(\mu^2 - m_V^2)^2 + \mu^2 \Gamma^2_{\rm tot}}\, \frac{\Gamma_{V \to \rm final\;state}}{\Gamma_{\rm tot}} ~.\label{inmedXsection}
\end{equation}
Here $\Gamma_{\rm tot}$ is the total width of the meson $V$, obtained as a sum
of the vacuum decay width, $\Gamma_{\rm vac}$, and an in-medium contribution $\Gamma_{\rm med}$:
\begin{equation}
\Gamma_{\rm tot} = \Gamma_{\rm vac} + \Gamma_{\rm med} ~
\end{equation}

with

\begin{equation}
\label{Gammamed}
\Gamma_{\rm med}(\rho(r)) = \Gamma_{\rm med}(\rho_0) \, \frac{\rho(r)}{\rho_0}
~. 
\end{equation}
in the low density approximation. Due to the second factor in Eq. \ref{inmedXsection} the meson decay into the channel of interest will decrease for a strong  broadening of the meson in the nuclear medium. Furthermore, according to the first factor in Eq. \ref{inmedXsection} this reduced yield is spread out over a broader mass range, making it much more difficult to separate the in-medium decay contribution from the background. Moreover, since $\Gamma_{\rm  tot}\sim \rho/\rho_0$ for $\Gamma_{\rm med} \gg \Gamma_{\rm vac}$ the second factor in Eq. \ref{inmedXsection} becomes proportional to $\frac{1}{\rho/\rho_0}$,  leading to a suppression of contributions from higher densities. The sensitivity of meson production in an elementary reaction is thereby shifted to the nuclear surface. In case of a strong in-medium broadening of a meson it is thus {\it in principle} difficult
to detect in-medium modifications by an analysis of the signal shape. As a consequence the
experiment becomes less sensitive to a possible mass shift. Requesting a
proton in coincidence with 3 photons does not shift the sensitivity to even
smaller densities. According to GiBUU simulations the fraction of $\omega
\rightarrow \pi^0 \gamma$ decays at densities larger than $0.1 \rho_0$ is
thereby changed only by less than 2\%  for the kinematic conditions of the
current analysis \cite{Friedrich}.\\
It should be pointed out, however, that a significant effect close to the
production threshold of the $\omega$ meson, $E_{\gamma}$=1109 MeV, was
nevertheless predicted by the Gi-BUU model ~\cite{gallmeister}. A data
analysis confined to this energy regime is under way and will be published separately.

\section{Summary and Conclusions}
Data on the photoproduction of $\omega$ mesons on $LH_2, C,$ and $ Nb$ have
been re-analyzed, applying an improved background determination and
subtraction method. An earlier claim of an in-medium lowering of the $\omega$
mass is not confirmed. The strong broadening of the $\omega$ meson in the
nuclear medium due to inelastic processes - as determined in a transparency
ratio measurement - suppresses contributions to the observed $\omega$ signal
from the interior of the nucleus. The branching ratio for in-medium decays
into the channel of interest is drastically reduced. Thereby, the sensitivity
is shifted to the nuclear surface, making the line shape analysis less
sensitive to a direct observation of in-medium modifications. Data with much
higher statistics will be needed to gain further insight. A corresponding
experiment has been performed at the MAMI C electron accelerator using the
Crystal Ball/TAPS detector setup. The analysis is ongoing.\\
\begin{acknowledgments}
We thank the scientific and technical staff at ELSA and the collaborating
institutions for their important contribution to the success of the
experiment. We acknowledge detailed discussions with M. Kaskulov, U. Mosel,
P. M\"uhlich, E. Oset and J. Weil. This work was supported financially by the {\it
  Deutsche Forschungs Gemeinschaft}  through SFB/TR16. The Basel group
acknowledges support from the {\it Schweizerischer Nationalfond} and the KVI
group from the {\it Stichting voor Fundamenteel Onderzoek der Materie (FOM)}
and the {\it Nederlandse Organisatie voor Wetenschappelijk Onderzoek (NWO)}.
\end{acknowledgments}

\bibliography{omega_Mariana_VM}

\end{document}